\documentclass[prb,twocolumn,showpacs,groupedaddress,citeautoscript]{revtex4}
\usepackage{graphicx}
\usepackage{amsmath,amssymb}
\graphicspath{{./img/}}
\newcommand{\T}{\mathcal{T}}
\renewcommand{\Re}{\mathop{\mathrm{Re}}}
\begin{document}
\title{Shot-noise control in ac-driven nanoscale conductors}
\author{S\'ebastien Camalet}
\altaffiliation[New address: ]{Laboratoire de Physique,
	Ecole normale sup\'erieure de Lyon,
	46, All\'ee d'Italie,
	69364 Lyon Cedex 07, France}
\affiliation{Institut f\"ur Physik, Universit\"at Augsburg,
        Universit\"atsstra\ss e~1, D-86135 Augsburg, Germany}
\author{Sigmund Kohler}
\affiliation{Institut f\"ur Physik, Universit\"at Augsburg,
        Universit\"atsstra\ss e~1, D-86135 Augsburg, Germany}
\author{Peter H\"anggi}
\affiliation{Institut f\"ur Physik, Universit\"at Augsburg,
        Universit\"atsstra\ss e~1, D-86135 Augsburg, Germany}
\date{\today}
%
\begin{abstract}

We derive within a time-dependent scattering formalism expressions for both
the current through ac-driven nanoscale conductors and its fluctuations.
The results for the time-dependent current, its time average, and, above
all, the driven shot noise properties assume an explicit and serviceable
form by relating the propagator to a non-Hermitian Floquet theory.
The driven noise cannot be expressed in terms of transmission probabilities.
The results are valid for a driving of arbitrary strength and frequency.
The connection with commonly known approximation schemes such as the
Tien-Gordon approach or a high-frequency approximation is elucidated
together with a discussion of the corresponding validity regimes.  Within
this formalism, we study the coherent suppression of current and noise
caused by properly chosen electromagnetic fields. 

\pacs{
05.60.Gg, 
85.65.+h, 
05.40.-a, 
72.40.+w  
}
\end{abstract}
\maketitle

\section{Introduction}

The experimental success in the coherent coupling of quantum dots
\cite{Blick1996a, Oosterkamp1998a, vanderWiel2003a} has enabled measuring
the transport properties of systems with a molecule-like level structure.
Recently, further progress in this direction has been attained by the
reproducible measurement of currents through molecules which are coupled to
metallic leads \cite{Cui2001a, Reichert2002a}.  Together with these
experimental achievements, new theoretical interest in the transport
properties of such nanoscale systems emerged \cite{Nitzan2001a,
Hanggi2002x}.  One particular field of interest is the interplay of the
electron transport and excitations by an oscillating gate voltage, a
microwave field, or an infrared laser, respectively.  Such excitations bear
intriguing phenomena like photon-assisted tunneling \cite{Tien1963a,
Inarrea1994a, Blick1995a, Stafford1996a, Brune1997a, Hazelzet2001a,
Kohler2002a, vanderWiel2003a, Platero2004a} and the adiabatic
\cite{Thouless1983a, Brouwer1998a, Altshuler1999a, Switkes1999a} and
non-adiabatic \cite{Wagner1999a, Levinson2000a, Wang2002a, Lehmann2002b}
pumping of electrons.

A prominent example for the control of quantum dynamics is the so-called
coherent destruction of tunneling, i.e., the suppression of the tunneling
dynamics of a particle in a double-well potential \cite{Grossmann1991a}, in
a two-level system \cite{Grossmann1992a}, or in a superlattice
\cite{Holthaus1992a}.  Recently, coherent destruction of tunneling has also
been found for the dynamics of two interacting electrons in a double
quantum dot \cite{Creffield2002a, Creffield2002b}.  Moreover, it has been
demonstrated that a corresponding transport effect exists: If two leads are
attached to the ends of a tunneling system, then a proper driving field can
be used to suppress the current even in the presence of a large transport
voltage \cite{Lehmann2003a}.  Moreover, in
such a system the corresponding shot noise level \textit{a priori} can be
controlled by proper ac fields \cite{Camalet2003a}.  Within this work, we
provide more details on this noise control scheme and also explore its
limitations.

An intuitive description of the electron transport through time-independent
mesoscopic systems is provided by the Landauer scattering formula
\cite{Landauer1957a} and its various generalizations.  Both the average
current \cite{Datta1995a} and the transport noise characteristics
\cite{Blanter2000a, Chen2003a} can be expressed in terms of the quantum
transmission coefficients for the respective scattering channels.  By
contrast, the theory for driven quantum transport is less developed.
Scattering of a single particle by an arbitrary time-dependent potentials
has been considered \cite{Henseler2000a, Henseler2001a, Li1999a} without
relating the resulting transmissions to a current between electron
reservoirs.  Such a relation is indeed non-trivial since the driving opens
inelastic transport channels and, therefore, in contrast to the static
case, an \textit{ad hoc} inclusion of the Pauli principle is no longer
unique.  This gave rise to a discussion about ``Pauli blocking factors''
\cite{Datta1992a, Wagner2000a}.  In order to avoid such conflicts, one
should start out from a many-particle description.  In this spirit, within
a Green function approach, a \textit{formal} solution for the current
through a time-dependent conductor has been presented, e.g., in
Refs.~~\onlinecite{Datta1992a} and ~\onlinecite{Jauho1994a} without taking
advantage of the full Floquet theory for the wire.  Nevertheless in some
special cases like, e.g., for conductors consisting of a single level
\cite{Wingreen1993a, Aguado1996a} or for the scattering by a piecewise
constant potential \cite{Inarrea1994a, Wagner1999a}, an explicit solution
becomes feasible.  Moreover, for large driving frequencies, the driving can
be treated within a self-consistent perturbation theory \cite{Brandes1997a,
Kohler2004a}.

The spectral density of the current fluctuations has been derived for the
low-frequency ac conductance \cite{Pretre1996a, Pedersen1998a} and the
scattering by a slowly time-dependent potential \cite{Lesovik1994a}.  For
arbitrary driving frequencies, the noise has been characterized by its
zero-frequency component \cite{Camalet2003a}.  A remarkable feature of the
current noise in the presence of time-dependent fields is its dependence on
the phase of the transmission \textit{amplitudes} \cite{Lesovik1994a,
Camalet2003a}.  By clear contrast, both the noise in the static case
\cite{Blanter2000a} and the current in the driven case \cite{Camalet2003a}
depend solely on transmission \textit{probabilities}.

Within this work, we \textit{derive} within a Floquet approach explicit
expressions for both the current and the noise properties of the electron
transport through a driven nanoscale conductor under the influence of
time-dependent forces.  This generalizes recent approaches since the
presented Floquet formalism is applicable to arbitrary periodically driven
tight-binding systems and, in particular, is valid for arbitrary driving
strength and, as well, extends beyond the adiabatic regime.  The dynamics of
the electrons is solved by integrating the Heisenberg equations of motion
for the electron creation/annihilation operators in terms of the
single-particle propagator.  For this propagator, in turn, we provide a
solution within a generalized Floquet approach.
Such a treatment is valid for effectively non-interacting electrons, i.e.,
when no strong correlations occur.  Disregarding these interactions also
implies that the displacement currents are not taken into account entirely.
As a consequence, the ac component of the electrical current
\textit{inside} the nanoconductor may deviate from the particle current
\cite{Landauer1992a, Blanter2000a}.

This paper is organized as follows.  After introducing in
Sec.~\ref{sec:model} a model for the leads and the conductor under the
influence of external fields, we derive in Sec.~\ref{sec:scattering} for a
situation with time-periodic but otherwise arbitrary driving general
expressions for the current and its noise and establish a connection to a
Floquet eigenvalue equation.  In Sec.~\ref{sec:approximations}, we consider
some special cases and approximations.  Section \ref{sec:application} is
devoted to the influence of an electromagnetic dipole field on a conductor
consisting of a few tight-binding levels.  Situations with an ac transport
voltage are addressed in Appendix~\ref{app:ac.voltage}, while in
Appendix~\ref{appendix:xi}, we detail an alternative derivation which has
been introduced in Ref.~~\onlinecite{Camalet2003a}.

\section{lead-wire model}
\label{sec:model}

\begin{figure}[tb]
\includegraphics[width=0.45\textwidth]{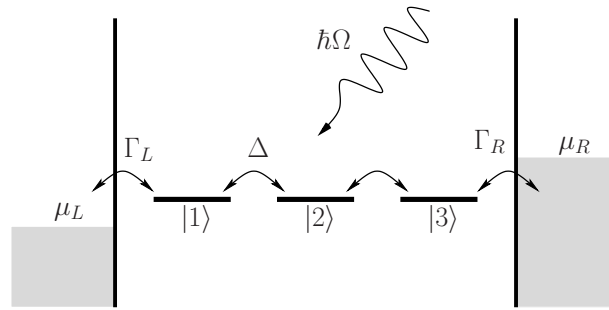}
\caption{\label{fig:levels} Level structure of a nano-conductor with
$N=3$ orbitals. The end sites are coupled to two leads with chemical
potentials $\mu_L$ and $\mu_R=\mu_L+eV$.}
\end{figure}

We start out by introducing a model for the central conductor (``wire'')
under the influence of an external driving field like, e.g., a molecular
wire subject to laser radiation or coupled quantum dots \cite{Blick1996a,
vanderWiel2003a} driven by microwaves or an oscillating gate voltage.  The
conductor is attached by tunneling couplings to external leads.  The entire
setup of our nanoscale system is described by the time-dependent
Hamiltonian
\begin{equation}
\label{eq:H}
H(t) = H_\mathrm{wire}(t) + H_{\rm leads} +
H_\mathrm{contacts},
\end{equation}
where the different terms correspond to the wire, the leads, and the
wire-lead couplings, respectively.  We focus on the regime of
coherent quantum transport where the main physics at work occurs on the
wire itself.  In doing so, we neglect other possible influences originating
from driving induced hot electrons in the leads, dissipation on the wire
and, as well, electron-electron interaction effects.  Then, the wire
Hamiltonian reads in a tight-binding approximation with $N$ orbitals
$|n\rangle$
\begin{equation}
H_{\rm wire}(t)= \sum_{n,n'} H_{nn'}(t) c^{\dag}_n c^{\phantom{\dag}}_{n'}\;.
\label{eq:Hw}
\end{equation}
For a molecular wire, this constitutes the so-called H\"uckel description
where each site corresponds to one atom.
The fermion operators $c_n$, $c_n^{\dag}$ annihilate and create,
respectively, an electron in the orbital $|n\rangle$.  The influence
of an applied ac field with frequency $\Omega=2\pi/{\cal T}$
results in a periodic time-dependence of the wire Hamiltonian:
$H_{nn'}(t+{\cal T})=H_{nn'}(t)$.  The leads are modeled by ideal
electron gases,
\begin{equation}
H_\mathrm{leads}=\sum_q \epsilon_{q} (c^{\dag}_{Lq}
c^{\phantom{\dag}}_{Lq} + c^{\dag}_{Rq} c^{\phantom{\dag}}_{Rq}),
\label{eq:Hl}
\end{equation}
where $c_{Lq}^{\dag}$ ($c_{Rq}^{\dag}$) creates an electron in the
state $|Lq \rangle$ ($|Rq \rangle$) in the left (right) lead.  The
tunneling Hamiltonian
\begin{equation}
H_{\rm contacts} = \sum_{q} \left( V_{Lq} c^{\dag}_{Lq} c^{\phantom{\dag}}_1
+ V_{Rq} c^{\dag}_{Rq} c^{\phantom{\dag}}_N
\right) + \mathrm{h.c.}
\label{eq:Hc}
\end{equation}
establishes the contact between the sites $|1\rangle$, $|N\rangle$
and the respective lead, as sketched in Fig.~\ref{fig:levels}.
This tunneling coupling is described by the spectral density
\begin{equation}
\label{eq:Gamma}
\Gamma_\ell (\epsilon) = 2\pi \sum_q |V_{\ell q}|^2
\delta(\epsilon-{\epsilon}_q)
\end{equation}
of lead $\ell$, $\ell=L,R$.  If the lead modes are dense,
$\Gamma_\ell(\epsilon)$ becomes a smooth function.

To fully specify the dynamics, we choose as an initial condition
for the left/right lead a grand-canonical electron ensemble at
temperature $T$ and electro-chemical potential $\mu_{L/R}$,
respectively.  Thus, the initial density matrix reads
\begin{equation}
\rho_0 \propto e^{-(H_\mathrm{leads} -\mu_L N_L -\mu_R N_R)/k_BT} ,
\label{eq:ic}
\end{equation}
where $N_{\ell}=\sum_q c^{\dag}_{\ell q} c^{\phantom{\dag}}_{\ell q}$
is the number of electrons in lead $\ell$
and $k_B T$ denotes the Boltzmann constant times temperature. 
An applied voltage $V$ maps to a chemical potential
difference $\mu_R-\mu_L=eV$ with $-e$ being the electron charge.
Then, at initial time $t_0$, the only nontrivial expectation values of the
wire operators read
$
\langle c_{\ell'q'}^\dagger c_{\ell q}\rangle
= f_\ell(\epsilon_q) \delta_{\ell\ell'}\delta_{qq'}
$
where $f_\ell(\epsilon)=(1+\exp[(\epsilon-\mu_\ell)/k_BT])^{-1}$ denotes
the Fermi function.

In our model Hamiltonian \eqref{eq:H}, the leads are time-independent.
Thus, it seemingly cannot describe ac transport voltages.  Such a
situation, however, can be mapped by a gauge transformation to one with
time-independent chemical potentials as demonstrated in
Appendix~\ref{app:ac.voltage}.

\section{scattering approach for time-dependent potentials}
\label{sec:scattering}

Due to their experimental accessibility, the central quantities in a
quantum transport problem are the stationary current and the low-frequency
part of its noise spectrum.  Within a scattering picture of
\textit{non-driven} mesoscopic transport, both quantities can be expressed
in terms of a transmission function $T(E)$ which reflects the probability
that an electron is transmitted from one lead to the other
\cite{Blanter2000a}.  Due to energy conservation, the reversed process,
occurs with equal probability.  This is no longer true for driven systems
and, consequently, the scattering approach needs to be generalized.  Thus,
in this section, we \textit{derive} expressions for the currents and its
noise properties for the transport through the time-dependent system
modeled above.  In the so-called wide-band limit, the more compact
derivation presented in Ref.~~\onlinecite{Camalet2003a} becomes possible,
cf.\ Appendix~\ref{appendix:xi}.  We will show that the average electrical
current contains only transition probabilities and, thus, resembles a
scattering formula.  In clear contrast to the static two-terminal case,
however, we will find that the noise depends in addition also on the phases
of the scattering matrix.

\subsection{Charge, current, and their fluctuations}

To avoid the explicit appearance of commutators in the definition of
correlation functions, we perform the derivation of the central transport
quantities in the Heisenberg picture.  As a starting point we choose the
operator
\begin{equation}
\label{eq:Q}
Q_\ell(t) = eN_\ell(t)-eN_\ell(t_0)
\end{equation}
that describes the charge accumulated in lead $\ell$ with respect to the
initial state.  Due to total charge conservation, $Q_\ell$ equals the net
charge transmitted across the contact $\ell$; its time derivative defines
the corresponding current
\begin{equation}
\label{eq:Ioperator}
I_\ell(t) = \frac{d}{dt} Q_\ell(t) .
\end{equation}
The current noise is described by the symmetrized correlation function
\begin{equation}
S_\ell (t,t')
= \frac{1}{2} \big\langle [\Delta I_\ell(t),\Delta I_\ell(t')]_+
\big\rangle
\label{eq:S}
\end{equation}
of the current fluctuation operator $\Delta I_\ell(t) = I_\ell(t)-\langle
I_\ell(t)\rangle$, where the anticommutator $[A,B]_+=AB+BA$ ensures
hermiticity.  It can be shown that at long times, $S_\ell(t,t') =
S_\ell(t+\mathcal{T},t'+\mathcal{T})$ shares the time-periodicity of the driving
\cite{on_periodicity}.  Therefore, it is possible to characterize the noise
level by the zero-frequency component of $S_\ell(t,t-\tau)$ averaged over the
driving period,
\begin{equation}
\label{eq:barS}
\bar S_\ell = \frac{1}{\cal T} \int_0^{\cal T} dt
\int_{-\infty}^\infty d\tau\, S_\ell (t,t-\tau) .
\end{equation}
We find below that for two-terminal devices $\bar S_\ell$ is
independent of the contact $\ell$, i.e., $\bar S_L=\bar S_R \equiv \bar S$.

The evaluation of the zero-frequency noise $\bar S$ directly from its
definition \eqref{eq:barS} can be tedious due to the explicit appearance of
both times, $t$ and $t-\tau$.  This inconvenience can be circumvented by
employing the relation
\begin{equation}
\label{eq:Qdiff}
\frac{d}{dt}\Big( \langle Q_\ell^2(t)\rangle - \langle Q_\ell(t)\rangle^2\Big)
= 2\int_0^\infty d\tau\, S_\ell(t,t-\tau)
\end{equation}
which follows from the integral representation of
Eqs.\ \eqref{eq:Q} and \eqref{eq:Ioperator},
$ Q_\ell(t) = \int_{t_0}^t dt'\, I_\ell(t')$, in the limit $t_0\to-\infty$.
By averaging Eq.~\eqref{eq:Qdiff} over the driving period and using
$S(t,t-\tau) = S(t-\tau,t)$, we obtain
\begin{equation}
\label{eq:barSQ}
\bar S = \Big\langle \frac{d}{dt}\langle\Delta
Q_\ell^2(t)\rangle\Big\rangle_t \,,
\end{equation}
where $\Delta Q_\ell = Q_\ell-\langle Q_\ell\rangle$ denotes the charge
fluctuation operator and $\langle\ldots\rangle_t$ the time average.
The fact that the time average can be evaluated from the limit $\bar S =
\lim_{t_0\to -\infty} \langle \Delta Q_\ell^2(t)\rangle/(t-t_0) > 0$ allows
to interpret the zero-frequency noise as the ``charge diffusion
coefficient''.
As a dimensionless measure for the \textit{relative} noise strength, we
employ the so-called Fano factor \cite{Fano1947a, on_factor2}
\begin{equation}
\label{eq:Fano}
F = \frac{{\bar S}}{e|\bar I|} \, ,
\end{equation}
where $\bar I$ denotes the time-average of the current expectation value
$\langle I_\ell(t)\rangle$.  Note that in a two-terminal device, the absolute
value of the average current is independent of the contact $\ell$.

\subsection{Transition amplitudes}

In order to take the exclusion principle properly into account,
we have formulated the transport problem under consideration in terms of
second quantization.  Nevertheless, in the absence of interactions, both the
current and its noise can be traced back to the solution of the corresponding
single-particle problem.  Thus, our next step is to relate the expectation
value and the variance of the charge operator \eqref{eq:Q} to the
transmission of electrons from one lead to the other.
For that purpose, we start from the Heisenberg equations of motion
\begin{align}
\label{eq:Heisenberg:lead}
\dot c_{L/Rq}
=& -\frac{i}{\hbar} \epsilon_q c_{L/Rq} - \frac{i}{\hbar}V_{L/Rq}\,c_{1/N} ,
\\
\label{eq:Heisenberg:1,N}
\dot c_{1/N}
=& -\frac{i}{\hbar} \sum_{n'} H_{1/N,n'}(t)\, c_{n'}
   -\frac{i}{\hbar} \sum_q V_{L/Rq}^* c_{L/Rq} ,
\\
\label{eq:Heisenberg:n}
\dot c_n =& -\frac{i}{\hbar} \sum_{n'} H_{nn'}(t)\, c_{n'}
\quad n=2,\ldots,N-1.
\end{align}
For these coupled linear equations, the formal solution
\begin{equation}
\begin{split}
c_{\ell' q'}(t)
={} & \sum_{\ell,q} \langle \ell' q' | U(t,t_0) | \ell q \rangle 
  \,c_{\ell q} (t_0) \\
  & + \sum_{n} \langle \ell' q' | U(t,t_0) | n \rangle \,c_{n}(t_0)
\end{split}
\label{eq:cLq}
\end{equation}
involves the propagator $U(t,t_0)$ of the corresponding single-particle
problem.  We insert \eqref{eq:cLq} into \eqref{eq:Q} and use the initial
condition \eqref{eq:ic} to obtain for the transfered charge at long times
[i.e., in the limit $t_0\to -\infty$, where all transients die out and, in
particular, the second line in Eq.~\eqref{eq:cLq} becomes irrelevant] the
expectation value
\begin{equation}
\label{eq:<Q>}
\langle Q_L(t)\rangle
= e\sum_{q',q,\ell}\left( |\langle L q'|U(t,t_0)|\ell q\rangle|^2
  - \delta_{\ell L}\delta_{qq'} \right) f_{\ell}(\epsilon_{q}) .
\end{equation}
To symmetrize this expression, we eliminate the backscattering terms,
i.e., the contributions with $\ell=L$, by employing the completeness relation
\begin{align}
\label{eq:completeness}
\mathbf{1} &=  \sum_q |Lq\rangle\langle Lq| + \sum_q |Rq\rangle\langle Rq| +
\sum_n |n\rangle\langle n|
\\
&\equiv  P_L + P_R +P_\mathrm{wire} ,
\end{align}
where $P_L$, $P_R$, and $P_\mathrm{wire}$ denote the projectors onto the
states of the left lead, the right lead, and the wire, respectively.  Then,
from the time derivative of Eq.~\eqref{eq:<Q>}, we find for the current
through the left contact the result
\begin{align}
\langle I_L(t) \rangle
={} & e \sum_{q,q'} \left\{ w_{Lq',Rq}(t) f_R(\epsilon_{q}) 
  - w_{Rq',Lq}(t) f_L(\epsilon_{q}) \right\} \nonumber \\
& -e \sum_{n,q} w_{n,Lq}(t) f_L(\epsilon_{q})
\label{eq:I_L(t)}
\end{align}
and \textit{mutatis mutandis} for the current through the right contact.
This expression already obeys the ``scattering form'' with the 
\textit{time-dependent} transmission
\begin{equation}
T_{\ell' \ell}(t,\epsilon)
= 2\pi\hbar \sum_{q,q'} w_{\ell'q',\ell q}(t)\, \delta(\epsilon-\epsilon_q)
\end{equation}
of electrons with energy $\epsilon$ from lead $\ell$ to lead $\ell'$.  At
asymptotic times, the transitions from the lead state $|\ell q\rangle$ to
the lead state $|\ell' q'\rangle$ and the wire state $|n\rangle$ happen
with the rates
\begin{align}
\label{eq:ratew}
w_{\ell' q',\ell q}(t)
=& \lim_{t_0 \to -\infty} \frac{d}{dt}
   \left| \big\langle \ell' q' | U(t,t_0) |\ell q \big\rangle \right|^2 ,
\\
w_{n,\ell q}(t)
=& \lim_{t_0 \to -\infty} \frac{d}{dt}
   \left| \big\langle n | U(t,t_0) |\ell q\big\rangle \right|^2 .
\end{align}
The last term in the current \eqref{eq:I_L(t)} describes a periodic
charging of the wire stemming from the external driving.  With an average
over one driving period, this contribution vanishes and, thus, the dc
current reads
\begin{equation}
\bar I
= e \sum_{q,q'} \left\{ \bar w_{Lq',Rq} f_R(\epsilon_{q}) 
  - \bar w_{Rq',Lq} f_L(\epsilon_{q}) \right\} ,
\label{eq:current}
\end{equation}
with $\bar w_{\ell' q',\ell q}$ denoting the time average of the rate
\eqref{eq:ratew}.  Interchanging in Eq.~\eqref{eq:current} $L$ and $R$
yields the negative current $-\bar I$.  Thus, as expected from total charge
conservation, the average current is, besides its sign, independent of the
contact at which it is evaluated.  We emphasize that \eqref{eq:current}
obeys the form of the current formula obtained for a \textit{static}
conductor within a scattering formalism.  In particular, consistent with
Refs.~~\onlinecite{Datta1995a} and ~\onlinecite{Datta1992a}, no ``Pauli
blocking factors'' $(1-f_\ell)$ appear in our derivation.  In contrast to a
static situation, this is in the present context relevant since for a
driven system generally $\bar w_{Lq,Rq'}\neq \bar w_{Rq',Lq}$, such that a
contribution proportional to $f_L(\epsilon_{q'}) f_R(\epsilon_q)$ would not
cancel \cite{Datta1992a, Wagner2000a}.

The zero-frequency noise $\bar S$ is conveniently derived from the charge
fluctuation with the help of relation \eqref{eq:barSQ}.  Expressing the
charge fluctuation by the Heisenberg operators \eqref{eq:cLq} yields for
the initial condition \eqref{eq:ic} after some algebra
\begin{equation}
\label{eq:Q2}
\begin{split}
\langle\Delta Q_L^2(t) \rangle \\
= \sum_{q,q'}\big\{
  & f_R(\epsilon_{q'})\bar f_R(\epsilon_{q})\,
    |\langle R{q'}|U^\dagger P_L U |Rq\rangle|^2 \\
+ & f_L(\epsilon_{q'})\bar f_R(\epsilon_{q})\,
    |\langle Lq'|U^\dagger P_L U |Rq\rangle|^2 \\
+ & f_L(\epsilon_{q'})\bar f_L(\epsilon_{q})\,
    |\langle L{q'}|U^\dagger (P_R+P_\mathrm{wire}) U |Lq\rangle|^2 \\
+ & f_R(\epsilon_{q'})\bar f_L(\epsilon_{q})\,
    |\langle Rq'|U^\dagger (P_R+P_\mathrm{wire}) U |Lq\rangle|^2
\big\} .
\end{split}
\end{equation}
By using the completeness relation \eqref{eq:completeness}, we have achieved a
form which is, besides the appearance of $P_\mathrm{wire}$, symmetric under
exchanging $L\leftrightarrow R$.  Here, $U$ is a shorthand notation for
$U(t,t_0)$ and $\bar f_\ell=1-f_\ell$.  Taking the time derivative and
averaging over the driving period yields
\begin{equation}
\begin{split}
{\bar S}
= e^2 \sum_{q,q'} \Big\{
  &  W^{L}_{Rq',Rq} \, f_R({\epsilon}_{q'}){\bar f}_R({\epsilon}_{q})
\\
+ &  W^{L}_{Lq',Rq} \, f_L({\epsilon}_{q'}){\bar f}_R({\epsilon}_{q}) 
\\
+ &  W^{R}_{Lq',Lq} \, f_L({\epsilon}_{q'}){\bar f}_L({\epsilon}_{q})
\\
+ &  W^{R}_{Rq',Lq} \, f_R({\epsilon}_{q'}){\bar f}_L({\epsilon}_{q}) 
\Big\} ,
\label{eq:Ssym}
\end{split}
\end{equation}
where we have defined
\begin{equation}
W_{\ell' q',\ell q}^{\ell''}
= \lim_{t_0\to -\infty} \Big\langle \frac{d}{dt}
  \big| \big\langle \ell' q' | U^{\dag}(t,t_0)
  P_{\ell''} U(t,t_0) |\ell q\big\rangle \big|^2 \Big\rangle_t .
\label{eq:W}
\end{equation} 
The contributions in Eq.~\eqref{eq:Q2} which contain the projector
$P_\mathrm{wire}$ on the wire states do not contribute to the
zero-frequency noise.
This can be demonstrated by inserting for the propagator the explicit
expressions \eqref{eq:UnlqFloquet} and \eqref{eq:UlqlqFloquet} which we
derive in the next subsection.
Interestingly enough, the noise $\bar S$ depends on both the diagonal and
the off-diagonal elements of the projector $U^\dagger P_{\ell''} U$.  By
contrast, the current \eqref{eq:I_L(t)} depends only on the diagonal
elements of this operator.  As a consequence, in the presence of driving it
is not possible to express the noise solely by transmission
\textit{probabilities}; cf.\ Eq.~\eqref{eq:Sfe}, below.

\subsection{Lead elimination}

The evaluation of the rates $w_{\ell'q',\ell q}$ and
$W^{\ell''}_{\ell'q',\ell q}$ involves the matrix elements of the
time-evolution operator $U(t,t_0)$ with the wire and the lead states.  In
the following, we eliminate the lead states and will find expressions for
the rates that depend explicitly only on the propagator for the wire
electrons and the spectral density of the couplings to the leads.

We start from the Schr\"odinger equation for the propagator,
$i\hbar\partial U(t,t')/\partial t = \mathcal{H}(t) U(t,t')$, where
$\mathcal{H}(t)$ is the single-particle Hamiltonian underlying
\eqref{eq:H}.  Formal integration with the initial condition
$U(t',t')=\mathbf{1}$ results in the Dyson equation
\begin{equation}
\label{eq:Dyson}
U(t,t') = U_0(t,t')-\frac{i}{\hbar}\int_{t'}^t dt''\, U_0(t,t'')
\mathcal{H}_\mathrm{contacts} U(t'',t'),
\end{equation}
where $U_0$ denotes the propagator in the absence of the wire-lead coupling.
We emphasize that due to the explicit time-dependence of the wire
Hamiltonian, the integral in \eqref{eq:Dyson} is not a mere convolution.
Using $\langle \ell'q'|U_0(t,t')|\ell q\rangle = \delta_{\ell\ell'}
\delta_{qq'}\exp[-i\epsilon_q(t-t')/\hbar]$, we find for the transition
matrix elements the relations
\begin{equation}
\label{eq:Unlq}
\langle n|U(t,t_0)|\ell q\rangle
= -\frac{i}{\hbar}V_{\ell q}^* \int_{t_0}^t dt'
e^{-\frac{i}{\hbar}\epsilon_q(t'-t_0)} 
   \langle n| U(t,t')| n_\ell\rangle
\end{equation}
and
\begin{multline}
\label{eq:Ulqlq}
\langle \ell'q'|  U(t,t_0) |\ell q\rangle \\
={}  e^{-\frac{i}{\hbar}(\epsilon_{q'}t-\epsilon_q t_0)}\Big\{
     \delta_{\ell\ell'}\delta_{qq'}
    -\frac{V_{\ell'q'}V_{\ell q}^*}{\hbar^2}
    \int_{t_0}^t\! dt' \int_{t_0}^{t'}\! dt''
\\  \times
    e^{\frac{i}{\hbar}(\epsilon_{q'} t' - \epsilon_q t'')}
    \langle n_{\ell'}| U(t',t'') | n_{\ell}\rangle \Big\} ,
\end{multline}
where $n_\ell$ denotes the wire site attached to lead $\ell$, i.e., $n_L=1$
and $n_R=N$.

At this stage, it is convenient to make use of the time-periodicity of the
Hamiltonian, $\mathcal{H}(t) = \mathcal{H}(t+\T)$.  This has the
consequence \cite{Grifoni1998a} that $U(t,t')=U(t+\T,t'+\T)$ and, thus, the
retarded Green function
\begin{equation}
\label{eq:Gtepsilon}
G(t,\epsilon)
=  -\frac{i}{\hbar} \int_0^\infty \! d\tau\, e^{\frac{i}{\hbar}\epsilon\tau}
    U(t,t-\tau)
= G(t+\mathcal{T},\epsilon)
\end{equation}
can be decomposed into a Fourier series,
$G(t,\epsilon) = \sum_{k=-\infty}^\infty e^{-ik\Omega t}
G^{(k)}(\epsilon)$, with the coefficients
\begin{align}
\label{eq:Gkepsilon}
G^{(k)}(\epsilon) = & \frac{1}{\mathcal{T}} \int_0^\mathcal{T} dt\,
e^{ik\Omega t} G(t,\epsilon) .
\end{align}
Physically, $G^{(k)}(\epsilon)$ describes the propagation of an electron
with initial energy $\epsilon$ under the absorption (emission) of $|k|$
photons for $k>0$ ($k<0$).  We emphasize that generally all sidebands
$k=-\infty\ldots\infty$ contribute to the Green function
\eqref{eq:Gtepsilon} and that, consequently, the $k$-summations are
unrestricted.

After making use of Eqs.~\eqref{eq:Gtepsilon} and \eqref{eq:Gkepsilon}, the
transition amplitudes \eqref{eq:Unlq} and \eqref{eq:Ulqlq} become
\begin{equation}
\label{eq:UnlqFloquet}
\langle n|U(t,t_0)|\ell q\rangle
= V_{\ell q}^* \, e^{-\frac{i}{\hbar}\epsilon_q(t-t_0)}
  \sum_k e^{-i k\Omega t} \langle n|G^{(k)}(\epsilon_q)|n_\ell\rangle
\end{equation}
and
\begin{equation}
\label{eq:UlqlqFloquet}
\begin{split}
\langle \ell'q'| U & (t,t_0) |\ell q\rangle \\
={} & e^{-\frac{i}{\hbar}(\epsilon_{q'}t-\epsilon_q t_0)}\Big\{
    \delta_{\ell\ell'}\delta_{qq'}
  - \sum_k V_{\ell'q'}V_{\ell q}^*
\\ & \times
    \frac{e^{\frac{i}{\hbar}
    (\epsilon_{q'} - \epsilon_q - k\hbar\Omega -i\eta)t}}
    {\epsilon_{q'} - \epsilon_q - k\hbar\Omega - i\eta}
    \langle n_{\ell'}| G^{(k)}(\epsilon_q) | n_{\ell}\rangle \Big\} ,
\end{split}
\end{equation}
respectively.
Since below we restrict ourselves to asymptotic times, $t_0\to -\infty$, we
have shifted the lower limit of the integrals accordingly.  Moreover, in
order to perform the $t'$-integration in Eq.\ \eqref{eq:Ulqlq}, we have
introduced a converging factor $e^{\eta t'/\hbar}$ and will finally
consider the limit $\eta\to 0$.

\begin{widetext}
\subsubsection{Average current}
\label{sec:barI}

For the further evaluation of the average current \eqref{eq:current}, we
insert the transition amplitude \eqref{eq:UlqlqFloquet} into
\eqref{eq:ratew}.  After taking the time derivative, averaging over time
$t$, and considering the limit $\eta\to 0$, we find
\begin{equation}
\label{eq:ratewFloquet}
\bar w_{Lq',Rq}
= \frac{2\pi}{\hbar} |V_{Lq'}V_{Rq}|^2
  \sum_k \big|G^{(k)}_{1N}(\epsilon_q)\big|^2
  \delta(\epsilon_{q'} - \epsilon_q - k\hbar\Omega) ,
\end{equation}
and the corresponding expression for $\bar w_{Rq',Lq}$. We have introduced
the notation $G_{nn'} = \langle n|G|n'\rangle$.
By use of the spectral density \eqref{eq:Gamma}, we replace the
remaining sums over the lead states by energy integrals and obtain as our
first main result the dc current
\begin{equation}
\label{eq:Ife}
\bar I
= \frac{e}{h}  \sum_{k=-\infty}^\infty \int d\epsilon
  \left\{ T_{LR}^{(k)} (\epsilon) f_R (\epsilon)
        - T_{RL}^{(k)} (\epsilon) f_L (\epsilon) \right\} ,
\end{equation}
where
\begin{align}
\label{eq:TLR}
T_{LR}^{(k)}(\epsilon)
=& \Gamma_L (\epsilon+k\hbar\Omega) \Gamma_R (\epsilon)
   \big|G_{1N}^{(k)}(\epsilon) \big|^2 , \\
\label{eq:TRL}
T_{RL}^{(k)}(\epsilon)
=& \Gamma_R (\epsilon+k\hbar\Omega) \Gamma_L (\epsilon)
   \big|G_{N1}^{(k)}(\epsilon) \big|^2 ,
\end{align}
denote the transmission probabilities for electrons from the right lead,
respectively from the left lead, with initial energy $\epsilon$ and final
energy $\epsilon+k\hbar\Omega$, i.e., the probability for an scattering
event under the absorption (emission) of $|k|$ photons if $k>0$ ($k<0$).

For a static situation, the transmissions $T_{LR}^{(k)}(\epsilon)$ and
$T_{RL}^{(k)}(\epsilon)$ are identical and contributions with $k\neq 0$
vanish.  Thus, it is possible to write the current \eqref{eq:Ife} as a
product of a \textit{single} transmission $T(\epsilon)$ and the difference
of the Fermi functions, $f_R(\epsilon)-f_L(\epsilon)$.  We emphasize that
in the driven case this is no longer true.

\subsubsection{ac current}

Although below we focus on the computation of dc currents, we here
continue the derivation of the transport quantities by presenting
explicit expressions for the ac currents.  We restrict ourselves to
$\langle I_L(t)\rangle$ since $\langle I_R(t)\rangle$ simply follows by
proper index replacements.  Evaluating $\langle I_L(t)\rangle$, we consider
also the last term in Eq.\ \eqref{eq:I_L(t)} which describes a periodic
charging/discharging of the wire.  Apart from the time average we perform
the same steps as in the derivation of the dc current and obtain
\begin{equation}
\label{eq:I(t)}
\langle I_L(t)\rangle
=  \frac{e}{h}  \int\! d\epsilon\,
   \big\{ T_{LR} (t,\epsilon) f_R (\epsilon)
        - T_{RL} (t,\epsilon) f_L (\epsilon) \big\}
   - \dot q_L(t)
\end{equation}
where
\begin{equation}
\begin{split}
q_{L}(t)
= & \frac{e}{2\pi} \int d\epsilon \, \Gamma_{L} (\epsilon)
  \sum_n \Big|\sum_{k} e^{-ik\Omega t}
     G_{n 1}^{(k)}(\epsilon) \Big|^2
    f_L (\epsilon)
\end{split}
\end{equation}
denotes the charge oscillating between the left lead and the wire.
Obviously, since $q_L(t)$ is time-periodic and bounded, its time derivative
cannot contribute to the average current.  The corresponding charge arising
from the right lead, $q_R(t)$, is \textit{a priori} unrelated to $q_L(t)$;
the actual charge on the wire reads $q_L(t)+q_R(t)$.  The time-dependent
current is determined by the time-dependent transmission
\begin{align}
\label{eq:tdT:LR}
T_{LR}(t,\epsilon)
=& \Gamma_R(\epsilon) \Re \sum_{k,k'} e^{-ik\Omega t} \,
   G_{1N}^{(k'+k)}(\epsilon) \big[ G_{1N}^{(k')}(\epsilon) \big]^* 
   \Big[ \Gamma_L (\epsilon+k'\hbar \Omega)
        +\frac{i}{\pi} \mathcal{P} \!\!\! \int d\epsilon' 
        \frac{\Gamma_L (\epsilon')}{\epsilon'-\epsilon-k'\hbar\Omega}
   \Big] .
\end{align}
The corresponding expression for $T_{RL}(t,\epsilon)$ follows from the
replacement $(L,1)\leftrightarrow(R,N)$.
Note that in the wide-band limit $\Gamma_\ell(\epsilon)=\Gamma_\ell$,
$\ell=L,R$, the contribution from the principal value integral vanishes.

\subsubsection{zero-frequency noise}

In order to obtain the zero-frequency noise $\bar S$, we evaluate the rates
$W^{\ell''}_{\ell'q',\ell q}$.  This step is performed along the lines of
reasoning for the evaluation of $\bar w_{\ell'q',\ell q}$ (although the
actual calculation is far more tedious): We insert the transition amplitude
\eqref{eq:UlqlqFloquet} into \eqref{eq:W}, take the derivative with
respect to $t$, and average over one driving period.
Finally, we employ the relation $\lim_{\eta\to 0} 4\eta [(\epsilon-a-i\eta)
(\epsilon-b+i\eta) (\epsilon'-b-i\eta) (\epsilon'-a+i\eta)]^{-1}
= (2\pi)^3 \delta(\epsilon-a)\, \delta(\epsilon'-b)\, \delta(a-b)$
to perform the limit $\eta\to 0$ and find
\begin{align}
\label{eq:W:RLR}
W_{Rq',Rq}^{L}
&=\frac{2 \pi}{\hbar} |V_{Rq'}|^2 |V_{Rq}|^2
  \sum_k
  \Big| \sum_{k'} \Gamma_L( \epsilon_{q}+k'\hbar\Omega) 
        \big[ G^{(k'-k)}_{1N}(\epsilon_{q'}) \big]^*
        G^{(k')}_{1N}({\epsilon}_{q})
  \Big|^2
  \delta(\epsilon_{q'} - \epsilon_q - k\hbar\Omega) ,
\\
\label{eq:W:LLR}
W_{Lq',Rq}^{L}
&=\frac{2 \pi}{\hbar} |V_{Lq'}|^2 |V_{Rq}|^2
   \sum_k
  \Big|
    \sum_{k'} \Gamma_L \big( \epsilon_{q}+k'\hbar\Omega \big) 
    \big[ G^{(k'-k)}_{11}({\epsilon}_{q'}) \big]^*
          G^{(k')}_{1N} ({\epsilon}_{q}) 
    -i G^{(k)}_{1N} ({\epsilon}_{q})
  \Big|^2 \delta(\epsilon_{q'}-{\epsilon}_{q}-k\hbar\Omega) .
\end{align}
The corresponding expressions for $W^R_{Lq',Lq}$ and $W^R_{Rq',Lq}$ follow
from the replacement $(L,1)\leftrightarrow(R,N)$.
Inserting these into the noise expression \eqref{eq:Ssym} we arrive at our
central result
\begin{equation}
\label{eq:Sfe}
\begin{split}
\bar S = \frac{e^2}{h} \sum_k \int d\epsilon \Big\{
  & \Gamma_R (\epsilon_k) \Gamma_R (\epsilon)
    \Big| \sum_{k'} \Gamma_L ( \epsilon_{k'})
    G^{(k'-k)}_{1N}(\epsilon_k) \big[G_{1N}^{(k')}(\epsilon)\big]^* \Big|^2
    f_R (\epsilon) \bar f_R (\epsilon_k) \\
+ & \Gamma_R (\epsilon_k) \Gamma_L (\epsilon) 
    \Big| \sum_{k'}
    \Gamma_L(\epsilon_{k'}) G^{(k'-k)}_{1N}(\epsilon_k)
    \big[ G_{11}^{(k')}(\epsilon) \big]^* -i G^{(-k)}_{1N}(\epsilon_k)
    \Big|^2 f_L(\epsilon) \bar f_R(\epsilon_k)
    \Big\} \\
+ & \text{same terms with the replacement $(L,1) \leftrightarrow (R,N)$} .
\end{split}
\end{equation}
We have defined $\epsilon_k=\epsilon+k\hbar\Omega$ and replaced the
sums over the lead states by energy integrations using the spectral density
\eqref{eq:Gamma}.
\end{widetext}

\subsection{Wide-band limit and Floquet theory}

In order to evaluate the expressions for $\bar I$ and $\bar S$ further, we
derive an eigenfunction representation for the Green function.  It is
well-known that beyond the adiabatic limit, the eigenfunctions of the
Hamiltonian are not of particular use --- rather a proper basis is provided
by a Floquet ansatz \cite{Shirley1965a, Sambe1973a, Grifoni1998a}.

Let us start from the Schr\"odinger equation for the propagator,
\begin{equation}
\label{eq:dotUnn}
i\hbar\frac{d}{dt} \langle n | U(t,t_0) | n' \rangle
= \sum_{n''} H_{nn''}(t) \langle n''|U(t,t_0)|n'\rangle ,
\end{equation}
for $n=2,\ldots,N-1$, and
\begin{align}
\label{eq:dotU1n}
i\hbar\frac{d}{dt} \langle n_\ell|U(t,t_0)|n'\rangle
=& \sum_{n''} H_{n_\ell n''}(t) \langle n''|U(t,t_0)|n'\rangle
   \nonumber \\
 & +\sum_q V_{\ell q}^* \langle \ell q| U(t,t_0) |n'\rangle ,
\end{align}
where $n_\ell$ is defined by $n_L=1$ and $n_R=N$.
To eliminate the lead states in the second line of Eq.~\eqref{eq:dotU1n},
we insert \eqref{eq:Unlq} and replace by use of the spectral density
\eqref{eq:Gamma} the sum over the lead states by an energy integral.
Then the last term in Eq.~\eqref{eq:dotU1n} becomes
\begin{equation}
\label{eq:Udiss}
-\frac{i}{2\pi\hbar}\int d\epsilon\, \Gamma_\ell(\epsilon) \int_{t_0}^t dt'
e^{-i\epsilon(t'-t_0)/\hbar} \langle n_\ell|U(t,t')|n'\rangle .
\end{equation}

Within the present context, we are mainly interested in the influence of
the driving field on the conductor and not in the details of the coupling
to the leads.  Therefore, we choose for $\Gamma_\ell(\epsilon)$ a rather
generic form by assuming that in the relevant regime, it is practically
energy-independent,
\begin{equation}
\Gamma_\ell(\epsilon) \longrightarrow \Gamma_\ell .
\end{equation}
This so-called wide-band limit allows further progress since we now can
perform in Eq.~\eqref{eq:Udiss} the remaining energy integration to obtain
$\hbar\delta(t'-t_0)$ and, consequently, Eq.~\eqref{eq:dotU1n} becomes
\begin{align}
\label{eq:dotU1nWBL}
i\hbar\frac{d}{dt} \langle n_\ell|U(t,t_0)|n'\rangle
=& \sum_{n''} H_{n_\ell n''}(t) \langle n''|U(t,t_0)|n'\rangle
   \nonumber \\
 & -\frac{i}{2}\Gamma_\ell \langle n_\ell|U(t,t_0)|n'\rangle .
\end{align}
Equations \eqref{eq:dotUnn} and \eqref{eq:dotU1nWBL}, together with the
initial conditions $\langle n|U(t,t)|n'\rangle=\delta_{nn'}$ fully
determine the propagator.  Solving this linear set of equations is
equivalent to computing a complete set of solutions for the equation
\begin{equation}
\label{eq:preFloquet}
i\hbar\frac{d}{dt} |\psi(t)\rangle
= \big(\mathcal{H}_\mathrm{wire}(t) - i\Sigma \big) |\psi(t)\rangle ,
\end{equation}
where the self-energy
\begin{equation}
\label{eq:sigma}
\Sigma = |1\rangle\frac{\Gamma_L}{2}\langle 1 |
       + |N\rangle\frac{\Gamma_R}{2}\langle N |
\end{equation}
results from the coupling to the leads.

Equation \eqref{eq:preFloquet} is linear and possesses time-dependent,
$\mathcal{T}$-periodic coefficients. Thus, it is possible to construct a
complete solution with the Floquet ansatz
\begin{align}
|\psi_\alpha(t)\rangle
= & \exp[(-i\epsilon_\alpha/\hbar-\gamma_\alpha)t] |u_\alpha(t)\rangle ,
\\
|u_{\alpha}(t)\rangle
= & \sum_{k=-\infty}^\infty |u_{\alpha,k}\rangle\exp(-ik\Omega t) .
\end{align}
The so-called Floquet states $|u_{\alpha}(t)\rangle$ obey the
time-periodicity of $\mathcal{H}_\mathrm{wire}(t)$ and have been decomposed
into a Fourier series.  In a Hilbert space that is extended by a periodic
time coordinate, the so-called Sambe space \cite{Sambe1973a}, they obey the
Floquet eigenvalue equation \cite{Grifoni1998a, Buchleitner2002a}
\begin{equation}
\Big(\mathcal{H}_\mathrm{wire}(t) - i\Sigma -i\hbar\frac{d}{dt}\Big)|u_{\alpha}(t)\rangle
= (\epsilon_{\alpha} -  i\hbar\gamma_{\alpha}) |u_{\alpha}(t)\rangle .
\label{eq:Fs}
\end{equation}
Due to the Brillouin zone structure of the Floquet spectrum
\cite{Shirley1965a, Sambe1973a, Grifoni1998a}, it is sufficient to compute
all eigenvalues of the first Brillouin zone,
$-\hbar\Omega/2<\epsilon_\alpha \le \hbar\Omega/2$.  Since the operator on
the l.h.s.\ of Eq.~\eqref{eq:Fs} is non-Hermitian, the eigenvalues
$\epsilon_{\alpha} - i\hbar\gamma_{\alpha}$ are generally complex valued
and the (right) eigenvectors are not mutually orthogonal.  Thus, to
determine the propagator, we need to solve also the adjoint Floquet
equation yielding again the same eigenvalues but providing the adjoint
eigenvectors $|u_\alpha^+(t)\rangle$.
It can be shown that the Floquet states $|u_\alpha(t)\rangle$ together with
the adjoint states $|u_\alpha^+(t)\rangle$ form at equal times a complete
bi-orthogonal basis: $\langle u^+_{\alpha}(t)|u_{\beta}(t)\rangle =
\delta_{\alpha\beta}$ and $\sum_{\alpha} |u_{\alpha}(t)\rangle \langle
u^+_{\alpha} (t)|= \mathbf{1}$.  A proof requires to account for the
time-periodicity of the Floquet states since the eigenvalue equation
\eqref{eq:Fs} holds in a Hilbert space extended by a periodic time
coordinate \cite{Jung1990a, Grifoni1998a}.

Using the Floquet equation \eqref{eq:Fs}, it is straightforward to show
that with the help of the Floquet states $|u_\alpha(t)\rangle$ the
propagator can be written as
\begin{equation}
\label{eq:Gtt}
U(t,t')
= \sum_\alpha e^{-i(\epsilon_\alpha/\hbar-i\gamma_\alpha)(t-t')}
  |u_\alpha(t)\rangle\langle u^+_\alpha(t')| ,
\end{equation}
where the sum runs over all Floquet states within one Brillouin zone.
Consequently, the Fourier coefficients of the Green
function [cf.\ Eq.\ \eqref{eq:Gkepsilon}] read
\begin{align}
G^{(k)}(\epsilon)
=& -\frac{i}{\hbar}\int_0^\T \frac{dt}{\T} e^{ik\Omega t}
   \int_0^\infty\!\! d\tau
   e^{i\epsilon\tau/\hbar} U(t,t-\tau)
\\
=& \sum_{\alpha}\sum_{k'=-\infty}^\infty
   \frac{|u_{\alpha,k'+k}\rangle\langle u_{\alpha,k'}^+|}
      {\epsilon-(\epsilon_\alpha+k'\hbar\Omega-i\hbar\gamma_\alpha)} .
\label{eq:G}
\end{align}
Inserting them into Eqs.\ \eqref{eq:Ife} and \eqref{eq:Sfe} yields
explicit expressions for the current and the noise, respectively.

\section{limiting cases}
\label{sec:approximations}

In the previous section, the dc current and the zero-frequency noise have
been derived for a periodic but otherwise arbitrary driving.
Within the wide-band limit, both quantities can be expressed in terms of
the solutions of the Floquet equation \eqref{eq:Fs}, i.e., the solution of
a non-Hermitian eigenvalue problem in an extended Hilbert space.  Thus, for
large systems, the numerical computation of the Floquet states can be
rather costly.  Moreover, for finite temperatures, the energy integration
in the expressions \eqref{eq:Ife} and \eqref{eq:Sfe} has to be performed
numerically.  Therefore, approximation schemes which allow a more efficient
computation are of much practical use.


Before introducing various approximation schemes for the wire propagator,
we discuss two particular cases for which current and noise assume more
intuitive expressions.  In doing so, we define quantities to which we will
refer later in this section.

\subsection{Static conductor and adiabatic limit}
\label{sec:static}

For consistency, the expressions \eqref{eq:Ife} and \eqref{eq:Sfe} for the
dc current and the zero-frequency noise, respectively, must coincide in the
undriven limit with the corresponding expressions of the time-independent
scattering theory \cite{Blanter2000a}.  This is indeed the case since the
static situation is characterized by two relations: First, in the absence
of spin-dependent interactions, we have time-reversal symmetry, $w_{Lq',Rq}
= w_{Rq,Lq'}$ and, second, all sidebands with $k\neq 0$ vanish, i.e.,
$T^{(k)}_{RL}(\epsilon) = T^{(k)}_{LR}(\epsilon) =
\delta_{k,0}T(\epsilon)$, where
\begin{equation}
\label{eq:Tstat}
T(\epsilon) = \Gamma_L(\epsilon)\, \Gamma_R(\epsilon)\, |G_{1N}(\epsilon)|^2
\end{equation}
and $G(\epsilon)$ is the Green function in the undriven limit.
Then the current assumes the known form
\begin{equation}
\label{eq:Istat}
I_0 = \frac{e}{h}\int d\epsilon\,T(\epsilon)
\big[ f_R(\epsilon)-f_L(\epsilon)\big] .
\end{equation}
Moreover in a static situation, the relation \cite{Datta1995a, on_datta}
\begin{equation}
\label{eq:datta}
| \Gamma_L(\epsilon) G_{11}(\epsilon) + i |^2 = 1-T(\epsilon) ,
\end{equation}
allows to eliminate the backscattering terms in the second line of
Eq.~\eqref{eq:Sfe} such that the zero-frequency noise can be expressed
solely in terms of the transmission to read \cite{Blanter2000a}
\begin{equation}
\label{eq:Sstat}
\begin{split}
S_0
= \frac{e^2}{h} \int \!\! d\epsilon \Big\{ &
  T(\epsilon) \left[ f_L ({\epsilon}) \bar f_L(\epsilon) + 
                     f_R ({\epsilon}) \bar f_R({\epsilon}) \right] 
\\
  + & T(\epsilon) \left[ 1-T(\epsilon) \right]  
   \left[ f_R (\epsilon) - f_L (\epsilon) \right]^2 \Big\} .
\end{split}
\end{equation}
For zero temperature, the terms in the first line vanish and pure shot
noise remains.  In contrast, for zero voltage, $f_R=f_L$ and the terms in
the first line constitute equilibrium quantum noise.  Obviously if both
voltage and temperature are zero, not only the current but also the noise
vanishes.  In the presence of driving, this is no longer the case.  This
becomes particularly evident in the high-frequency limit studied in
Sec.~\ref{sec:approx:hf}.

It is known that in the adiabatic limit, i.e., for small driving frequencies,
the numerical solution of the Floquet equation \eqref{eq:Fs} becomes
infeasible because a diverging number of sidebands has to be taken into
account.  In more mathematical terms, Floquet theory has no proper limit as
$\Omega\to 0$ \cite{Hone1997a}.  The practical consequence of this is that for
low driving frequencies, it is favorable to tackle the transport problem with
a different strategy: If $\hbar\Omega$ is the smallest energy-scale of the
Hamiltonian \eqref{eq:H}, one computes for the ``frozen'' Hamiltonian at each
instance of time the current and the noise from the static expressions
\eqref{eq:Istat} and \eqref{eq:Sstat} being followed up by time-averaging.

\subsection{Infinite voltage}

Many phenomena can be discussed in the limit
of very large (practically infinite; subscript $\infty$) voltages such that
$f_R \rightarrow 1$ and $f_L \rightarrow 0$ in the relevant energy range.
Then, the dc current \eqref{eq:Ife} becomes
\begin{equation}
\label{eq:Iinfgen}
\bar I_\infty = \frac{e}{h} \sum_k \int d\epsilon\,
\Gamma_L(\epsilon+k\hbar\Omega) \Gamma_R(\epsilon)
|G_{1N}^{(k)}(\epsilon)|^2 .
\end{equation}
In the zero-frequency noise \eqref{eq:Sfe}, only the contribution with $f_R
\bar f_L$ remains, thus,
$\bar S_\infty = e^2 \sum_{q',q} W_{Rq',Lq}^{R}
= e^2 \sum_{q,q'} ( w_{Rq',Lq} - W_{Lq',Lq}^{R} )$.
To derive this expression, we again have used the completeness relation
\eqref{eq:completeness} and the fact that terms containing the projector on
the wire states do not contribute to time averages.  Expressing
$w_{Rq',Lq}$ and $W_{Lq',Lq}^{R}$ by the Green functions yields
\begin{equation}
\label{eq:Sinfgen}
\begin{split}
\bar S_\infty
=  e \bar I_\infty 
   - & \frac{e^2}{h} \sum_k \int\! d\epsilon \,
   \Gamma_L (\epsilon_k) \Gamma_L (\epsilon) 
\\ \times & 
   \Big| \sum_{k'} \Gamma_R ( \epsilon_{k'} ) G^{(k'-k)}_{N1} ({\epsilon}_{k}) 
   \big[G_{N1}^{(k')} ({\epsilon})\big]^* \Big|^2 ,
\end{split}
\end{equation}
where $\epsilon_k=\epsilon+k\hbar \Omega$. 
These expressions make explicit that $\bar I_\infty>0$ and $\bar S_\infty <
e\bar I_\infty$.  Consequently, for infinite voltage the Fano factor
\eqref{eq:Fano} cannot exceed unity.

\subsection{Weak wire-lead coupling}
\label{sec:approx:weakcoupl}

In the limit of a weak wire-lead coupling, i.e., for coupling constants
$\Gamma_\ell$ which are far lower than all other energy scales of the wire
Hamiltonian, it is possible to derive within a master equation approach
a closed expression for the dc current \cite{Lehmann2003b}.  The
corresponding approximation within the present Floquet approach is based on
treating the self-energy contribution $-i\Sigma$ in the non-Hermitian
Floquet equation \eqref{eq:Fs} as a perturbation.  Then, the zeroth order
of the Floquet equation
\begin{equation}
\Big(\mathcal{H}_\mathrm{wire}(t)-i\hbar\frac{d}{dt}\Big)|\phi_{\alpha}(t)\rangle
= \epsilon^0_{\alpha} |\phi_{\alpha}(t)\rangle ,
\end{equation}
describes the driven wire in the absence of the leads,
where $|\phi_\alpha(t)\rangle=\sum_k \exp(-ik\Omega t) |\phi_{\alpha,k}
\rangle$ are the ``usual'' Floquet states with quasienergies
${\epsilon}_{\alpha}^0$.  In the absence of degeneracies the first
order correction to the quasienergies is $-i\hbar\gamma_\alpha^1$ where
\begin{align}
\gamma_\alpha^1
={}& \frac{1}{\hbar}\int_0^{\mathcal{T}} \frac{dt}{\mathcal{T}}
   \langle\phi_\alpha(t)| \Sigma |\phi_\alpha(t)\rangle
\\
={}& \frac{\Gamma_L}{2\hbar} \sum_k |\langle 1|\phi_{\alpha,k}\rangle|^2
+  \frac{\Gamma_R}{2\hbar} \sum_k |\langle N|\phi_{\alpha,k}\rangle|^2 .
\label{eq:gamma_alpha}
\end{align}
Since the first-order correction to the Floquet states
will contribute to neither the current nor the noise, the zeroth-order
contribution $|u_\alpha(t)\rangle = |u_\alpha^+(t)\rangle =
|\phi_\alpha(t)\rangle $ is already sufficient for the present purpose.
Consequently, the transmission \eqref{eq:TLR} assumes the form
\begin{equation}
\label{eq:Tweak}
\begin{split}
T^{(k)}_{LR}(\epsilon)
= \Gamma_L\Gamma_R \sum_{\alpha,\beta,k',k''} &
\frac{\langle N|\phi_{\alpha,k'}\rangle\langle\phi_{\alpha,k'+k}|1\rangle}
     {\epsilon-(\epsilon_\alpha^0+k'\hbar\Omega+i\hbar\gamma_\alpha^1)}
\\ \times &
\frac{\langle 1|\phi_{\beta,k''+k}\rangle\langle\phi_{\beta,k''}|N\rangle}
     {\epsilon-(\epsilon_\beta^0+k''\hbar\Omega-i\hbar\gamma_\beta^1)}
\end{split}
\end{equation}
and $T_{RL}^{(k)}(\epsilon)$ accordingly.  The transmission
\eqref{eq:Tweak} exhibits for small values of $\Gamma_\ell$ sharp peaks at
energies $\epsilon^0_\alpha+k'\hbar\Omega$ and
$\epsilon^0_\beta+k''\hbar\Omega$ with widths $\hbar\gamma_\alpha^1$ and
$\hbar\gamma_\beta^1$.  Therefore, the relevant contributions to the sum
come from terms for which the peaks of both factors coincide and, in the
absence of degeneracies in the quasienergy spectrum, we keep only terms
with
\begin{equation}
\alpha=\beta,\quad k'=k'' .
\end{equation}
Then, the fraction in \eqref{eq:Tweak} is a Lorentzian and can be
approximated by
$\pi\delta(\epsilon-\epsilon_\alpha^0-k'\hbar\Omega)/\hbar\gamma_\alpha^1$
provided that $\gamma_\alpha^1$ is small.  Consequently, the energy
integration in \eqref{eq:Ife} can be performed even for finite temperature
and we obtain for the dc current the expression
\begin{equation}
\label{eq:I:RWA}
\bar I
= \frac{e}{\hbar}\sum_{\alpha,k,k'}  
    \frac{\Gamma_{L\alpha k} \Gamma_{R\alpha k'}}
     {\Gamma_{L\alpha}+\Gamma_{R\alpha}}
   \big[ f_R(\epsilon_{\alpha}^0+k'\hbar\Omega)-
                f_L(\epsilon_{\alpha}^0+k\hbar\Omega)\big] .
\end{equation}
The coefficients
\begin{align}
\Gamma_{L\alpha k} =& \Gamma_L|\langle 1|\phi_{\alpha,k}\rangle|^2, &
 \Gamma_{L\alpha} =& \sum_k \Gamma_{L\alpha k} \,, \\
\Gamma_{R\alpha k} =& \Gamma_R|\langle N|\phi_{\alpha,k}\rangle|^2 , &
 \Gamma_{R\alpha} =& \sum_k \Gamma_{R\alpha k} \,,
\end{align}
denote the overlap of the $k$th sideband $|\phi_{\alpha,k}\rangle$ of
the Floquet state $|\phi_\alpha(t)\rangle$ with the first site and the last
site of the wire, respectively.  We have used $2\hbar\gamma_\alpha^1 =
\Gamma_{L\alpha} + \Gamma_{R\alpha}$ which follows from
\eqref{eq:gamma_alpha}.  Expression \eqref{eq:I:RWA} has been derived in a
prior work \cite{Lehmann2003b} within a rotating-wave approximation of a
Floquet master equation approach.

Within the same approximation, we expand the
zero-frequency noise \eqref{eq:Sfe} to lowest order in $\Gamma_{\ell}$:
After inserting the spectral representation \eqref{eq:G} of the Green
function, we again keep only terms with identical Floquet index $\alpha$ and
identical sideband index $k$ to obtain
\begin{equation}
\label{eq:S:RWA}
\begin{split}
\bar S
= & \frac{e^2}{\hbar} \sum_{\alpha,k,k'}
\frac{\Gamma_{R\alpha k'}\bar f_R(\epsilon_\alpha^0{+}k'\hbar\Omega)}
     {(\Gamma_{L\alpha}+\Gamma_{R\alpha})^3}
\big\{
  2\Gamma_{L\alpha}^2 \Gamma_{R\alpha k}f_R(\epsilon_\alpha^0{+}k\hbar\Omega)
\\ & \hspace{17ex}
+ (\Gamma_{L\alpha}^2 + \Gamma_{R\alpha}^2)
  \Gamma_{L\alpha k} f_L(\epsilon_\alpha^0 {+} k\hbar\Omega)
\big\}
\\ &
+ \,\text{same terms with the replacement $L \leftrightarrow R$} .
\end{split}
\end{equation}

Of particular interest for the comparison to the static situation is the
limit of a large applied voltage such that practically $f_R=1$ and $f_L=0$.
Then, in Eqs.\ \eqref{eq:I:RWA} and \eqref{eq:S:RWA}, the sums over the
sideband indices $k$ can be carried out such that
\begin{align}
\label{eq:IinfG0}
\bar I_\infty
=& \frac{e}{\hbar} \sum_{\alpha} 
\frac{\Gamma_{L\alpha} \Gamma_{R\alpha}}{\Gamma_{L\alpha}+\Gamma_{R\alpha}}
, \\
\label{eq:SinfG0}
\bar S_\infty
=& \frac{e^2}{\hbar} \sum_{\alpha} 
   \frac{\Gamma_{L\alpha}\Gamma_{R\alpha}(\Gamma_{L\alpha}^2
   +\Gamma_{R\alpha}^2)}{(\Gamma_{L\alpha}+\Gamma_{R\alpha})^3} .
\end{align} 
These expressions resemble the corresponding expressions for the transport
across a \textit{static} double barrier \cite{Blanter2000a}.  If now
$\Gamma_{L\alpha}=\Gamma_{R\alpha}$ for all Floquet states
$|\phi_\alpha(t)\rangle$, we find $F=1/2$.  This is in particular the case
for systems obeying reflection symmetry \cite{on_symmetry}.
\nocite{Lehmann2003b} In the presence of such symmetries,
however, the existence of exact crossings, i.e.\ degeneracies, limits the
applicability of the weak-coupling approximation.

\subsection{Homogeneous ac driving}
\label{sec:approx:homo}

In many experimental situations, the driving field acts as a time-dependent
gate voltage, i.e., it merely shifts all on-site energies of the wire
uniformly.  Thus, the wire Hamiltonian is of the form
\begin{equation}
\mathcal{H}_\mathrm{wire}(t)=\mathcal{H}_0 + f(t)\sum_n |n \rangle \langle n| ,
\end{equation}
where, without loss of generality, we restrict $f(t)$ to possess zero
time-average.  A particular case of such a homogeneous driving is realized
with a system that consists of only one level \cite{Wingreen1990a,Kislov1991a,
Aguado1996a}.  Then trivially, the time and the position dependence of the
Floquet states factorize and, therefore, the dc current can be obtained
within the formalism introduced by Tien and Gordon \cite{Tien1963a}.  Here, we establish the relation between
such a treatment and the present Floquet approach.

Since the time-dependent part of the Hamiltonian is proportional to the
unity operator, the solution of the Floquet equation \eqref{eq:Fs} is,
besides a phase factor, given by the eigenfunctions $|\alpha\rangle$ of
the static operator $\mathcal{H}_0-i\Sigma$,
\begin{equation}
\label{eq:states:tucker}
|u_\alpha(t)\rangle = e^{-i F(t)} |\alpha\rangle ,
\end{equation}
where $(\mathcal{H}_0-i\Sigma)|\alpha\rangle =
(\epsilon_\alpha-i\hbar\gamma_\alpha)|\alpha\rangle$ and
$dF(t)/dt=f(t)/\hbar$.
The quasienergies $(\epsilon_\alpha-i\hbar\gamma_\alpha)$ coincide with the
eigenvalues of the static eigenvalue problem.  Note that $F(t)$ obeys the
$\T$-periodicity of the driving field since the time-average of $f(t)$
vanishes by definition.  Thus, the phase factor in the Floquet states
\eqref{eq:states:tucker} can be written as a Fourier series,
\begin{equation}
e^{-i F(t)} = \sum_k a_k\, e^{-ik\Omega t}
\end{equation}
and, consequently we find $|u_{\alpha,k}\rangle=a_k|\alpha\rangle$ and the
adjoint states accordingly.
Then, the Green function \eqref{eq:Gkepsilon} becomes
\begin{equation}
\label{eq:GTucker}
G^{(k)}(\epsilon)
= \sum_{k'}  a_{k'+k}\, a_{k'}^*\, G(\epsilon-k'\hbar\Omega) ,
\end{equation}
where $G(\epsilon)$ denotes the Green function in the absence of the
driving field.  Inserting \eqref{eq:GTucker} into \eqref{eq:Ife} and
employing the sum rule $\sum_{k'} a_{k'}^* a_{k'+k}=\delta_{k,0}$, yields
\begin{equation}
\bar I = \sum_k |a_k|^2
\frac{e}{h} \int\! d\epsilon \, T (\epsilon-k\hbar\Omega)
 [ f_{R}(\epsilon) - f_{L}(\epsilon) ] ,
\label{eq:ITucker}
\end{equation}
where $T(\epsilon)$ is the transmission in the absence of the driving.
This expression allows the interpretation, that for homogeneous driving,
the Floquet channels contribute \textit{independently} to the current
${\bar I}$.  For the special case of a one-site conductor and a sinusoidal
driving, this relation to the static situation has been discussed in
Refs.~~\onlinecite{Wingreen1990a} and ~\onlinecite{Kislov1991a}.

Addressing the noise properties, we obtain by inserting the Green function
\eqref{eq:GTucker} into \eqref{eq:Sfe} the expression
\begin{widetext}
\begin{equation}
\begin{split}
\label{eq:STucker}
\bar S = \frac{e^2}{h}\sum_k \int\! d\epsilon\, \Big\{ &
\Big| \sum_{k'} a_{k'+k}^* a_{k'} T(\epsilon-k'\hbar\Omega) \Big|^2
f_R(\epsilon) \bar f_R(\epsilon+k\hbar\Omega)
\\
+ & \Gamma_L \Gamma_R
\Big| \sum_{k'} a_{k'+k}^* a_{k'} G_{1N}(\epsilon-k'\hbar\Omega)
\big[ \Gamma_L G_{11}^*(\epsilon-k'\hbar\Omega) - i \big] \Big|^2
f_L(\epsilon) \bar f_R(\epsilon+k\hbar\Omega)
\\
+ & \,\text{same terms with the replacement $(L,1) \leftrightarrow (R,N)$}
\Big\} .
\end{split}
\end{equation}
\end{widetext}
While the term in the first line contains only the static transmission at
energies shifted by multiples of the photon energies, the contribution in
the second line cannot be brought into such a convenient form.  The reason
for this is that the sum over $k'$ inhibits the application of the relation
\eqref{eq:datta}.  As a consequence, in clear contrast to the dc current,
the zero-frequency noise cannot be interpreted in terms of independent
Floquet channels.
Only in the limit of large driving frequencies, we find below that the
channels become effectively independent and \eqref{eq:STucker} reduces to
an expression that depends only on the transmission in the absence of the
driving and the Fourier coefficients $a_k$, cf.\ next subsection.

Expressions for the dc current and the noise that depend only on the static
transmission have been derived by Tucker and Feldman \cite{Tucker1979a,
Tucker1985a} within a Tien-Gordon approach \cite{Tien1963a}.  The central
approximation of this approach is the description of a time-dependent
chemical potential by an effective electron distribution.  While this
yields the correct expression \eqref{eq:ITucker} for the dc current, it does
not capture the interference terms in the noise formula \eqref{eq:STucker}.
This reveals that a Tien-Gordon-like approach yields the correct dc current
while for the noise (and other higher-order correlation functions) it is
only valid in a high-frequency limit.

For large voltages where $f_L=0$ and $f_R=1$, the sums over the Fourier
coefficients in Eqs.\ \eqref{eq:ITucker} and \eqref{eq:STucker} can be
evaluated with the help of the sum rule $\sum_{k'} a_{k'}^*
a_{k'+k}=\delta_{k,0}$.  Then both the dc current and the zero-frequency
noise become identical to their value in the absence of the driving.  This
means that for a sufficiently large transport voltage, a time-dependent
gate voltage has no influence on the average current and the zero-frequency
noise.

\subsection{High-frequency driving}
\label{sec:approx:hf}

Many effects occurring in driven quantum systems, such as coherent
destruction of tunneling \cite{Grossmann1991a} or current and noise control
\cite{Lehmann2003a, Camalet2003a}, are most pronounced for large excitation
frequencies.  Thus, it is particularly interesting to derive for the
present Floquet approach an expansion in terms of $1/\Omega$.  Thereby, the
driven system will be approximated by a static system with renormalized
parameters.  Such a perturbation scheme has been developed for two-level
systems in Ref.~~\onlinecite{Shirley1965a} and applied to driven tunneling
in bistable systems \cite{Grossmann1991b} and superlattices
\cite{Holthaus1992a}.  For open quantum system, the coupling to the
external degrees of freedom (e.g., the leads or a heat bath) bears
additional complications that have been solved heuristically in
Ref.~~\onlinecite{Kohler2004a} by replacing the Fermi functions by
effective electron distributions.  In the following, we present a rigorous
derivation of this approach based on a perturbation theory for the Floquet
equation~\eqref{eq:Fs}.

We assume a driving that leaves all off-diagonal matrix elements of the
wire Hamiltonian time-independent while the tight-binding levels undergo a
position-dependent, time-periodic driving $f_n(t)=f_n(t+\T)$ with zero
time-average.  Then, the wire Hamiltonian is of the form
\begin{equation}
\label{eq:Mel}
\mathcal{H}_\mathrm{wire}(t)
=\mathcal{H}_0 + \sum_n f_n(t)\, |n \rangle \langle n| .
\end{equation}
If $\hbar\Omega$ represents the largest energy scale of the problem, we can
in the Floquet equation \eqref{eq:Fs} treat the \textit{static} part of the
Hamiltonian as a perturbation.  Correspondingly, the eigenfunctions of the
operator $\sum_n f_n(t)|n\rangle\langle n| -i\hbar d/dt$ determine the zeroth
order Floquet states
\begin{equation}
e^{-i F_n (t)}|n\rangle .
\label{eq:zthorder}
\end{equation}
We have defined the function
\begin{equation}
\label{eq:Fn}
F_n(t) = \frac{1}{\hbar} \int_0^t dt'\, f_n(t') = F_n(t+\T),
\end{equation}
which is $\T$-periodic due to the zero time-average of $f_n(t)$.  As a
consequence of this periodicity, to zeroth order the quasienergies are zero
($\mathrm{mod}\,\hbar\Omega$) and the Floquet spectrum is
given by multiples of the photon energy, $k\hbar\Omega$.  Each
$k=0,\pm1,\pm2,\ldots$ defines a degenerate subspace of the extended
Hilbert space.  If now $\hbar\Omega$ is larger than all other energy
scales, the first-order correction to the Floquet states and the
quasienergies can be calculated by diagonalizing the perturbation in the
subspace defined by $k=0$.  Thus, we have to solve the time-independent
eigenvalue equation
\begin{equation}
(\mathcal{H}_\mathrm{eff}-i\Sigma)|\alpha\rangle
= (\epsilon^{1}_{\alpha}-i\hbar \gamma^{1}_{\alpha}) |\alpha\rangle .
\label{eq:Hzthorder}
\end{equation}
The time-independent effective Hamiltonian $\mathcal{H}_\mathrm{eff}$ is
defined by the matrix elements of the original static Hamiltonian
$\mathcal{H}_0$ with the zeroth order Floquet states \eqref{eq:zthorder},
\begin{equation}
\label{eq:Heff}
(\mathcal{H}_\mathrm{eff})_{nn'}
= \int_0^\T \frac{dt}{\T}
  e^{iF_n(t)} (\mathcal{H}_0)_{nn'} e^{-iF_{n'}(t)} .
\end{equation}
The $t$-integration constitutes the inner product in the Hilbert space
extended by a periodic time coordinate \cite{Sambe1973a}.
To first order in $1/\Omega$, the quasienergies $\epsilon^1_\alpha-i\hbar
\gamma^1_\alpha$ are given by the eigenvalues of the static equation
\eqref{eq:Hzthorder} and, consequently, the corresponding Floquet states read
\begin{equation}
\label{eq:hf:states}
|u_\alpha(t)\rangle = \sum_n e^{-iF_n(t)}|n\rangle\langle n|\alpha\rangle .
\end{equation}
The fact that all $F_n(t)$ are $\T$-periodic, allows to write in
\eqref{eq:hf:states} the time-dependent phase factor as a Fourier series,
\begin{equation}
\label{eq:a:hf}
e^{-iF_n(t)} = \sum_k a_{n,k}\,e^{-ik\Omega t} .
\end{equation}
Thus, $\langle n|u_{\alpha,k}\rangle = a_{n,k}\langle n|\alpha\rangle$ and
the Green function for the high-frequency driving reads
\begin{equation}
\label{eq:UlO}
G_{nn'}^{(k)}(\epsilon) = \sum_{k'} a_{n,k'+k} a_{n',k'}^*
G_{nn'}^\mathrm{eff}(\epsilon-k'\hbar\Omega) ,
\end{equation}
where $G^\mathrm{eff}(\epsilon)$ denotes the Green function corresponding
to the static Hamiltonian $\mathcal{H}_\mathrm{eff}$ with the self-energy
$\Sigma$.  Finally, substituting $\epsilon\to\epsilon+k'\hbar\Omega$ and
using the sum rule $\sum_{k'} a_{n,k+k'} a_{n,k'}^* = \delta_{k,0}$, we obtain
\begin{equation}
\bar I = \frac{e}{h} \int d\epsilon \; T_\mathrm{eff}(\epsilon)
\big\{ f_{R,\mathrm{eff}}(\epsilon) - f_{L,\mathrm{eff}}(\epsilon) \big\} .
\label{eq:IlO}
\end{equation}
The effective transmission $T_\mathrm{eff}(\epsilon) = \Gamma_L \Gamma_R
|G_{1N}^\mathrm{eff}(\epsilon)|^2$ is computed from the effective
Hamiltonian \eqref{eq:Heff}; the electron distribution is given by
\begin{align}
f_{L,\mathrm{eff}}(\epsilon)
= \sum_k |a_{1,k}|^2 f_L(\epsilon+k\hbar\Omega)
\label{eq:feff}
\end{align}
and $f_{R,\mathrm{eff}}$ follows from the replacement $(1,L)\to(N,R)$.

In order to derive a high-frequency approximation for the zero-frequency
noise $\bar S$, we insert the Green function \eqref{eq:UlO} into
\eqref{eq:Sfe} and neglect products of the type
$G^\mathrm{eff}(\epsilon-k\hbar\Omega)\,
G^\mathrm{eff}(\epsilon-k'\hbar\Omega)$ for $k\neq k'$.  Employing the
above sum rule for the Fourier coefficients $a_{n,k}$, we obtain for the
noise the static expression \eqref{eq:Sstat} but with the transmission
$T(\epsilon)$ and the Fermi functions $f_{R,L}(\epsilon)$ replaced by the
effective transmission $T_\mathrm{eff}(\epsilon)$ and the effective
distribution function \eqref{eq:feff}, respectively.  The fact that
$f_{L,\mathrm{eff}}(\epsilon)$ is generally not a mere Heaviside step
function has an intriguing consequence: In the presence of driving, the
noise remains finite even if both voltage and temperature are zero.

Two differences between the high-frequency approximation and the
homogeneous driving, cf.\ Sec.~\ref{sec:approx:homo}, are worth mentioning:
First, the static transmission is now replaced by an effective transmission
which can be considerably influenced by the driving, see below.
Second, in general $a_{1,k}\neq a_{N,k}$ such that
$f_{R,\mathrm{eff}}\neq f_{L,\mathrm{eff}}$.  This means that the driving
can create an effective bias and thereby create a non-adiabatic pump
current.  By contrast Eq.\ \eqref{eq:ITucker} reveals that a homogeneous
driving cannot create such a pump current.
Moreover, if all $F_n$ are identical as in the case of a homogeneous
driving, the effective Hamiltonian $\mathcal{H}_\mathrm{eff}$ equals the
original static Hamiltonian.  Then, also the second line of
Eq.~\eqref{eq:STucker} can be written in terms of the static transmission
$T(\epsilon)$.

\section{Conductor driven by an oscillating dipole field}
\label{sec:application}

In this section, we apply the formalism derived in
Secs.~\ref{sec:scattering} and \ref{sec:approximations} to study the
conduction and noise properties of a nanoscale conductor under the
influence of an electromagnetic field.  As an elementary model that
captures the essential features of a molecular wire \cite{Nitzan2001a}, we
employ a tight-binding model composed of $N$ sites as sketched in
Fig.~\ref{fig:levels}.  Each orbital is coupled to its nearest neighbor by
a hopping matrix element $\Delta$, thus, the single-particle wire
Hamiltonian reads
\begin{equation}
\label{eq:Hdipole}
\begin{split}
\mathcal{H}_\mathrm{wire}(t)
= & -\Delta\sum_{n=1}^{N-1}\big( |n\rangle\langle n{+}1|
    + |n{+}1\rangle\langle n| \big) \\
& + \sum_n [E_n + f_n(t)]\,|n\rangle\langle n| ,
\end{split}
\end{equation}
where $E_n$ denote the on-site energies of the tight-binding levels.
Within a dipole approximation, the oscillating electromagnetic field causes
the time-dependent level shifts
\begin{equation}
f_n(t)=A \cos (\Omega t) x_n
\label{eq:acdf}
\end{equation}
with $x_n=(N+1-2n)/2$ the scaled position of site $|n\rangle$.  Since
typical laser frequencies are below the work function of a usual metal, we
assume that the radiation does not penetrate the leads and that,
consequently, the leads stay in thermal equilibrium.  The energy $A$
denotes the electrical field amplitude multiplied by the electron charge
and the distance between two neighboring sites and, thus, depends
implicitly on the length of the sample.  This model describes, as well, an
array of coherently coupled quantum dots \cite{Blick1996a, Oosterkamp1998a,
vanderWiel2003a} under the influence of microwave radiation.

The dipole approximation inherent to the driving~\eqref{eq:acdf} neglects
the propagation of the electromagnetic field and, thus, is valid only for
wavelengths that are much larger than the size of the sample
\cite{Pellegrini1993a}.  This condition is indeed fulfilled for both
applications we have in mind: For molecular wires, we consider frequencies
up the optical spectral range, i.e., wavelengths of the order
$1\,\mu\mathrm{m}$ and samples that extend over a few nanometers.  Coupled
quantum dots typically \cite{Blick1996a, Oosterkamp1998a, vanderWiel2003a}
have a distance of less than $1\,\mu\mathrm{m}$ while the coupling matrix
element $\Delta$ is of the order of $30\,\mu\mathrm{eV}$ which corresponds
to a wavelength of roughly $1\,\mathrm{cm}$.

We assume that the wire couples equally strong to both leads, thus,
$\Gamma_L=\Gamma_R\equiv\Gamma$.  An applied transport voltage $V$ is
mapped to a symmetric shift of the leads' chemical potentials, $\mu_R =
-\mu_L = eV/2$.  Moreover, for the evaluation of the dc current and the
zero-frequency noise, we restrict ourselves to zero temperature.  The
zero-temperature limit is physically well justified for molecular wires at
room temperature and for quantum dots at helium temperature since in both
cases thermal electron excitations do not play a significant role.

\subsection{Current and noise suppression}

For a wire described by the Hamiltonian \eqref{eq:Hdipole}, it has been
found \cite{Camalet2003a} that a dipole force of the form \eqref{eq:acdf}
suppresses the transport if the ratio $A/\hbar\Omega$ is close to a zero of
the Bessel function $J_0$ (i.e., values 2.405.., 5.520.., 8.654.., \ldots).
Moreover, in the vicinity of such suppressions, the shot noise
characterized by the Fano factor \eqref{eq:Fano} assumes two characteristic
minima.  These suppression effects are most pronounced in the
high-frequency regime, i.e., if the energy quanta $\hbar\Omega$ of the
driving exceed the energy scales of the wire.  Thus, before going into a
detailed discussion, we start with a qualitative description of the effect
based on the static approximation for a high-frequency driving that has
been derived in Sec.~\ref{sec:approx:hf}.

Let us consider first the limit of a voltage which is so large that in Eq.\
\eqref{eq:IlO}, $f_{R,\mathrm{eff}} - f_{L,\mathrm{eff}}$ can be replaced by
unity.  Then, the average current is determined by the effective
Hamiltonian
\begin{equation}
\mathcal{H}_\mathrm{eff}
=  -\Delta_\mathrm{eff}\sum_{n=1}^{N-1}\big( |n\rangle\langle n{+}1|
    + |n{+}1\rangle\langle n| \big)
 + \sum_{n=1}^N E_n \,|n\rangle\langle n| ,
\end{equation}
which has been derived by inserting the driving \eqref{eq:acdf} into Eqs.\
\eqref{eq:Fn} and \eqref{eq:Heff}.  Then,
obviously $\mathcal{H}_\mathrm{eff}$ is identical to the Hamiltonian
\eqref{eq:Hdipole} in the absence of the driving field but with the tunnel
matrix element renormalized according to
\begin{equation}
\Delta \to \Delta_\mathrm{eff} = J_0 (A/\hbar\Omega) \Delta .
\label{eq:effhop}
\end{equation}
Since the Bessel function $J_0$ assumes values between zero and one, the
amplitude of the driving field allows to switch the absolute value of the
effective hopping on the wire, $\Delta_\mathrm{eff}$, between 0 and
$\Delta$.  Since the transmission of an undriven wire is proportional to
$|\Delta|^2$, the effective transmission $T_\mathrm{eff}(\epsilon)$
acquires a factor $J_0^2(A/\hbar\Omega)$.  This renormalization of the
hopping results finally in a current
suppression\cite{Lehmann2003a,Camalet2003a}.

For the discussion of the shot noise, we employ the Fano factor
\eqref{eq:Fano} as a measure.  In the limit of large applied voltages,
we have to distinguish two limits:
(i) weak wire-lead coupling $\Gamma\ll\Delta_\mathrm{eff}$ (i.e., weak
with respect to the effective hopping) and
(ii) strong wire-lead coupling $\Gamma\gg\Delta_\mathrm{eff}$.
In the first case, the tunnel contacts between the lead and the wire act as
``bottlenecks'' for the transport.  In that sense they form barriers.  Thus
qualitatively, we face a double barrier situation and, consequently,
expect the shot noise to exhibit a Fano factor $F\approx1/2$
\cite{Blanter2000a}.  In the second case, the links between the wire sites
act as $N-1$ barriers.  Correspondingly, the Fano factor assumes values
$F\approx 1$ for $N=2$ (single barrier) and $F\approx 1/2$ for $N=3$
(double barrier) \cite{on_multibarrier}\nocite{DeJong1995a}.  At the
crossover between the two limits, the conductor is optimally ``barrier
free'' such that the Fano factor assumes its minimum.

In order to be more quantitative, we evaluate the current and the
zero-frequency noise in more detail thereby considering a finite voltage.
This requires a closer look at the effective electron distribution
\eqref{eq:feff}; in particular, we have to quantify the concept of a
``practically infinite'' voltage.  In a static situation, the voltage can
be replaced by infinity, $f_R(\epsilon) = 1 = 1-f_L(\epsilon)$, if all
eigenenergies of the wire lie well inside the range $[\mu_L,\mu_R]$.  In
contrast to the Fermi functions, the effective electron distribution
\eqref{eq:feff} which is decisive here, decays over a broad range in
multiple steps of size $\hbar\Omega$.  Since for our model,
$T_\mathrm{eff}(\epsilon)$ is peaked around $\epsilon=0$, we replace here
the effective electron distributions by their values for $\epsilon=0$,
\begin{equation}
\label{eq:feff:K(V)}
f_{\ell,\mathrm{eff}}(0)
= \sum_{k<\mu_{\ell}/\hbar\Omega} J_k^2\Big(\frac{A(N-1)}{2\hbar\Omega}\Big) ,
\end{equation}
for zero temperature.  We have inserted the coefficients $a_{1,k} =
J_k(A(N-1)/2\hbar\Omega)$ and $a_{N,k} = J_{-k}(A(N-1)/2\hbar\Omega)$
which have been computed directly from their definition \eqref{eq:a:hf};
$J_k$ denotes the $k$th order Bessel functions of the first kind.
The current, the noise, and the Fano factor are given by the static
expressions \eqref{eq:Istat} and \eqref{eq:Sstat} with the transmission and
the electron distribution replaced by the corresponding effective
quantities, $T_\mathrm{eff}$ and $f_{\mathrm{eff},\ell}$, respectively.
Thus, we obtain
\begin{align}
\label{eq:Ihf}
\bar I =& \lambda \bar I_{\infty} ,
\\
\label{eq:Shf}
\bar S =& \lambda^2 \bar S_{\infty}
          + \frac{e}{2}(1-\lambda^2) \bar I_{\infty} ,
\\
F =& \lambda F_\infty + \frac{1-\lambda^2}{2\lambda},
\label{eq:Fhf}
\end{align}
respectively, where the subscript $\infty$ denotes the corresponding
quantities in the infinite voltage limit,
\begin{align}
\bar I_\infty =& \frac{e}{h}\int d\epsilon\, T_\mathrm{eff}(\epsilon) ,
\\
\bar S_\infty =& \frac{e^2}{h}\int d\epsilon\, T_\mathrm{eff}(\epsilon)
            [1 - T_\mathrm{eff}(\epsilon)] ,
\end{align}
and $F_\infty=\bar S_\infty/e\bar I_\infty$.
The factor
\begin{equation}
\label{eq:lambda}
\lambda
= f_{R,\mathrm{eff}}(0) - f_{L,\mathrm{eff}}(0)
= \sum_{|k|\leq K(V)} J_k^2 \Big(\frac{A(N-1)}{2\hbar\Omega}\Big)
\end{equation}
reflects the influence of a finite voltage; $K(V)$ denotes the
largest integer not exceeding $e|V|/2\hbar\Omega$.
Since $J_k(x)\approx 0$ for $|k|>x$ and $\sum_k J_k^2(x)\approx 1$, we find
$\lambda=1$ if $K(V)>A(N-1)/2\hbar\Omega$.  This means that for small
driving amplitudes $A<eV/(N-1)$, we can consider the voltage as practically
infinite.  With an increasing driving strength, $\lambda$ decreases and,
thus, the current becomes smaller by a factor $\lambda$ but still exhibits
suppressions.  By contrast, since $F_\infty\leq 1$ for all situations
considered here [cf.\ the remark after Eq.~\eqref{eq:Sinfgen}], we find
from Eq.~\eqref{eq:Fhf} that the Fano factor will increase with smaller
$\lambda$.

\subsection{Numerical results}

The qualitative discussion of the current and noise suppressions can be
corroborated by exact numerical results.  For this purpose, we have solved
numerically the Floquet equation \eqref{eq:Fs}.  With the resulting Floquet
states and quasienergies, we obtained the Green function \eqref{eq:G}.  In
the zero temperature limit considered here, the Fermi functions in the
expressions for the average current \eqref{eq:Ife} and the zero-frequency
noise \eqref{eq:Sfe} become step functions.  The remaining energy
integrals can be performed analytically since the integrands are rational
functions.

\subsubsection{Intermediate wire-lead coupling}

\begin{figure}[tb]
\includegraphics{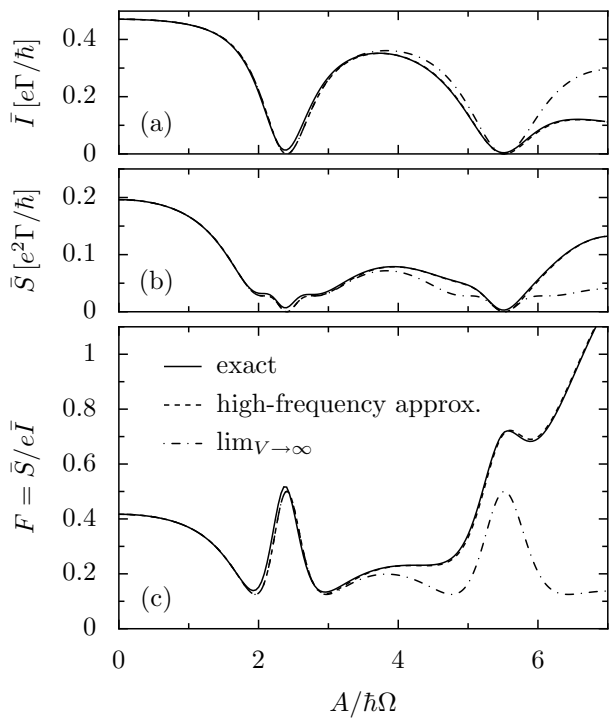}
\caption{\label{fig:hfr} Time-averaged current (a), zero-frequency noise
(b), and Fano factor (c) for a conductor consisting of $N=3$ sites with
equal on-site energies, $E_n=0$, as a functions of the driving amplitude
$A$.  The driving frequency is $\Omega=5\Delta/\hbar$, the wire-lead
coupling is $\Gamma=0.5\Delta$, and the chemical potentials are
$\mu_R=-\mu_L=24\Delta$.  The exact numerical results (solid lines) are
compared to the high-frequency approximation for finite (dashed) and
infinite voltage (dash-dotted).}
\end{figure}

Figure~\ref{fig:hfr} depicts the average current, the zero-frequency
noise, and the corresponding Fano factor for a wire that consists of $N=3$
sites with on-site energies $E_n=0$ as sketched in Fig.~\ref{fig:levels}.
The driving frequency $\Omega=5\Delta/\hbar$ lies above all transition
energies of the wire states and the applied voltage $V=48\Delta/e$ is
relatively large.  This particular value of the voltage has been selected
to avoid chemical potentials to lie close to multiples of $\hbar\Omega$,
i.e., close to the steps of the effective electron distribution
\eqref{eq:feff:K(V)}.  The wire-lead coupling $\Gamma=0.5\Delta$ is
sufficiently weak, such that in the absence of the driving, the transport
is dominated by resonant tunneling.  Correspondingly, the current is
essentially determined by the hopping rate $\Gamma/2\hbar$ of the electrons from
the lead to the wire.  The noise exhibits a Fano factor $F\approx 1/2$
which is the characteristic value for the transport across a double barrier
\cite{DeJong1995a,Blanter2000a}.  With an increasing driving amplitude, the
current becomes smaller until it reaches its minimum when the ratio
$A/\hbar\Omega$ assumes a zero of the Bessel function $J_0$.  Note that
while the analytical treatment within a high-frequency approximation
predicts a vanishing current, the exact result is still roughly 1\% of the
value in the absence of the driving.  Close to the current suppression, the
effective tunnel matrix element \eqref{eq:effhop} is much smaller than the
wire-lead coupling $\Gamma$ and the connections to the central site of the
wire form a double barrier.  Consequently, we again find a Fano factor
$F\approx 1/2$.  At the crossover $\Delta_\mathrm{eff} \approx \Gamma$, the
effective barriers vanish and, therefore, the Fano factor assumes its
minimum.  These exact numerical results are well reproduced by the
expressions \eqref{eq:Ihf}--\eqref{eq:Fhf} obtained within a
high-frequency approximations for finite voltage.  Figure \ref{fig:hfr}
also reveals that for small driving amplitudes, $A<eV/(N-1)$, the
assumption of a practically infinite voltage yields the correct results.  By
contrast, for larger driving amplitudes, $A>eV/(N-1)$, the Fano factor can
assume values $F>1$, i.e., the shot noise becomes even larger than in the
static situation.
 
As the coupling strength $\Gamma$ is lowered, the distance between a pair
of minima of the Fano factor becomes smaller until the minima finally
vanish \cite{Camalet2003a} (not shown).  In this limit, the current and the
noise are given by the weak coupling results \eqref{eq:IinfG0} and
\eqref{eq:SinfG0}, respectively.  The corresponding Fano factor $F=1/2$
[cf.\ the discussion after Eq.\ \eqref{eq:SinfG0}] is independent of the
driving amplitude.

\begin{figure}[tb]
\includegraphics{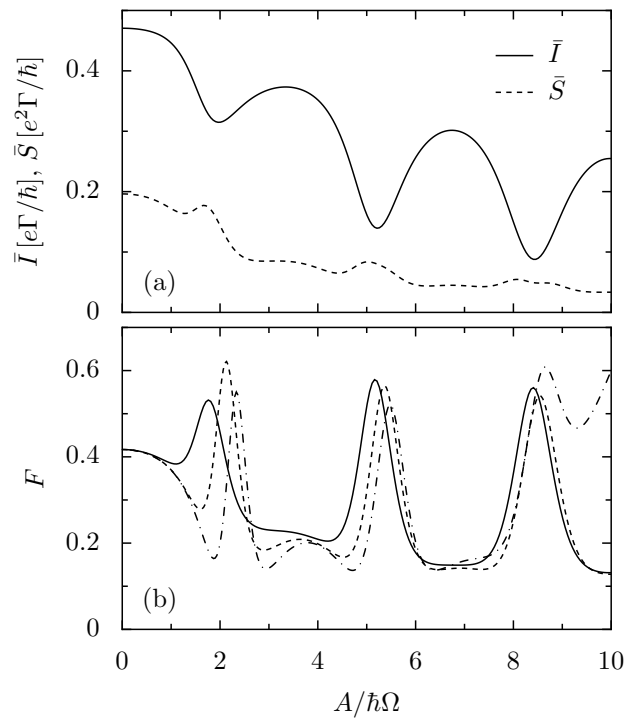}
\caption{\label{fig:lf}
(a) Time-averaged current (solid line) and zero-frequency noise (dashed
line) as a function of the driving amplitude for the driving frequency
$\Omega=\Delta/\hbar$.
(b) Corresponding Fano factor for the same data (solid line) and for the
driving frequencies $\Omega=1.5\Delta/\hbar$ (broken) and
$\Omega=3\Delta/\hbar$ (dash-dotted).
All other parameters are as in Fig.~\ref{fig:hfr}.
}
\end{figure}

Figure~\ref{fig:lf}a depicts the behavior for a driving frequency which is
of the order of the wire excitations, $\Omega=\Delta/\hbar$.  Then, the
high-frequency approximation is no longer applicable.  Nevertheless, the
average current exhibits clear minima with a reduction of the order 50\%.
Compared to the high-frequency case, these minima are shifted towards
smaller driving amplitudes, i.e., they occur for ratios $A/\hbar\Omega$
slightly below the zeros of the Bessel function $J_0$.  At the minima of
the current, the Fano factor (solid line in Fig.~\ref{fig:lf}b) still
assumes a maximum with a value close to $F\approx 1/2$.  Although, the
sharp minima close to the current suppressions have vanished, in-between
the maxima the Fano factor assumes remarkably low values ($F\approx 0.2$).
Figure~\ref{fig:lf}b also reveals that already for $\Omega\approx
3\Delta/\hbar$, the high-frequency regime is reached.

\subsubsection{Strong wire-lead coupling}

\begin{figure}[tb]
\includegraphics{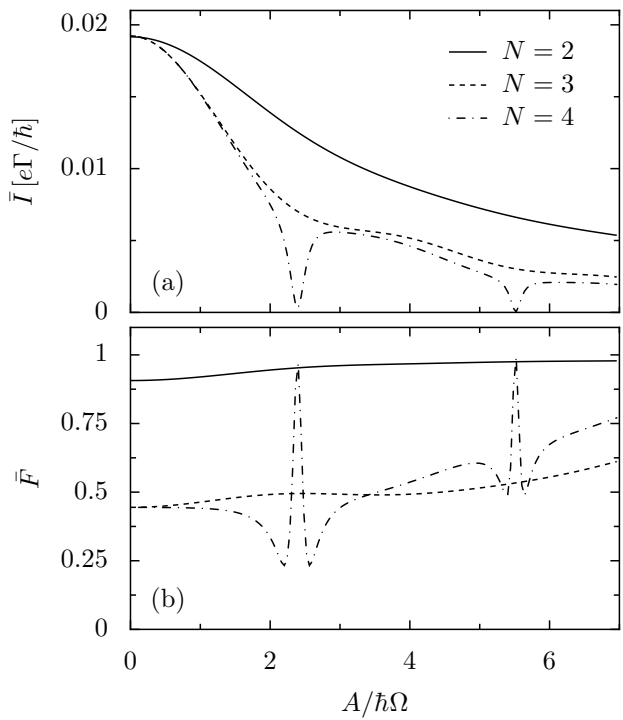}
\caption{\label{fig:lG} Time-averaged current (a) and Fano factor (b) as a
function of the driving amplitude $A$ for a wire with $N=2,3,4$ sites and
the wire-lead coupling strength $\Gamma = 10\Delta$. The other parameters
are $E_n=0$, $\Omega=5\Delta/\hbar$ and $\mu_R=-\mu_L=25\Delta$.}
\end{figure}

For strong wire-lead coupling, it is possible to choose a driving frequency
that is large with respect to the wire excitations, but small as compared
to the coupling $\Gamma$, thus $\Delta \ll \hbar\Omega \ll \Gamma$.
Figure~\ref{fig:lG} depicts the current and the Fano factor in this limit
for wires with a different number of sites.  The qualitative difference
between these cases can be explained by the fact that due to the strong
coupling, the first and the last wire site hybridize with the leads.  Then
the setup behaves similar to a wire with $N-2$ sites and a weak wire-lead
coupling $\propto \Delta^2/\Gamma$.  This means that for $N=2$ the wire
acts as point contact while for $N=3$, we qualitatively have resonant
transport through a single level.  In both cases remains no tunneling
matrix element of the wire that could be renormalized and, consequently,
for $N\leq 3$ the current suppressions vanish in the strong-coupling limit
(cf.\ Fig.~\ref{fig:lG}a).  This scenario is also reflected in the behavior
of the Fano factor (Fig.~\ref{fig:lG}b) which exhibits the characteristic
values $F\approx 1$ (point contact) for $N=2$ and $F\approx 1/2$ (single
resonant level) for $N=3$.  Finally, for $N=4$ we observe the behavior of a
driven wire with two sites and weak coupling \cite{Kohler2004a}.  Then, a
vanishing effective hopping $\Delta_\mathrm{eff}\approx 0$ corresponds to a
point contact, thus, $F\approx 1$.  Although the behavior of the Fano
factor can be explained by drawing analogies to a weakly coupled wire with
$N-2$ sites, the global decay of the current with the driving amplitude,
cf.\ Fig.~\ref{fig:lG}a, is not within the scope of this intuitive picture.

\subsubsection{Internal bias}

\begin{figure}[tb]
\includegraphics{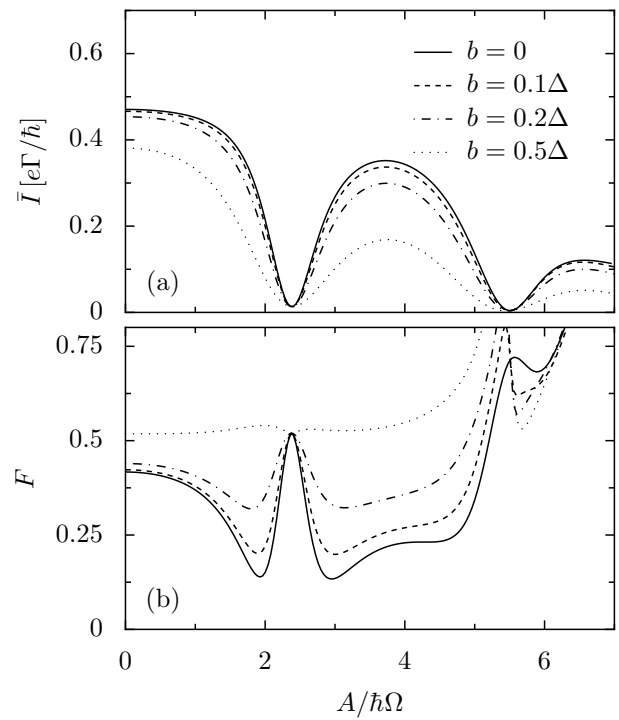}
\caption{\label{fig:bias} Time-averaged current (a) and Fano factor (b) as
a function of the driving amplitude $A$ for a wire with $N=3$ sites in the
presence of an internal bias.  The on-site energies are $E_1=b$, $E_2=0$,
$E_3=-b$.  All other parameters are as in Fig.~\ref{fig:hfr}.}
\end{figure}

So far, we have assumed that all on-site energies of the wire are
identical.  In an experimental setup, however, the applied transport
voltage acts also a static dipole force which rearranges the charge
distribution in the conductor and thereby causes an internal potential
profile \cite{Nitzan2002a, Pleutin2003a, Liang2004a}.  The self-consistent
treatment of such effects is, in particular in the time-dependent case,
rather ambitious and beyond the scope of this work.  Thus, here we only
derive the consequences of a static bias without determining its shape from
microscopic considerations.  We assume a position-dependent static shift of
the on-site energies by an energy $-b\,x_n$, i.e., for a wire with $N=3$
sites,
\begin{equation}
\label{eq:Ebias}
E_1=b,\quad E_2=0,\quad E_3=-b .
\end{equation}
Figure \ref{fig:bias}a demonstrates that the behavior of the average
current is fairly stable against the bias.  In particular, we still find
pronounced current suppressions.  Note that since $b\ll\Omega$ a
high-frequency approximation is still possible.  As a main effect of the
bias, we find reduced current maxima while the minima remain.  By
contrast, the minima of the Fano factor (Fig.~\ref{fig:bias}b) become
washed out:  Once the bias becomes of the order of the wire-lead coupling,
$b\approx\Gamma$, the structure in the Fano factor vanishes and we find
$F\approx 1/2$ for all driving amplitudes $A<eV/(N-1)$ [cf.\ the discussion
discussion after Eq.~\eqref{eq:lambda}].  Interestingly, the value of the
Fano factor at current suppressions is bias independent.

\section{Conclusions}

We have derived with Eqs.\ \eqref{eq:Ife}--\eqref{eq:TRL}, \eqref{eq:I(t)}, and
\eqref{eq:Sfe} expressions for the dc current, the zero-frequency noise,
and the time-dependent current for the electron transport through ac-driven
nanoscale systems.  A cornerstone of our approach is the relation of the
propagator to a non-Hermitian Floquet equation.  This yields explicit
formulae for the current and the noise.  Moreover, the connection to
Floquet theory allows to elucidate various approximation schemes that enable an
efficient computation and, in addition, provide physical insight.
Above all, a high-frequency approximation has emerged to be very useful: Within
an expansion in $1/\Omega$, the driven transport problem can
be approximated by a time-independent transport problem with a renormalized
tunneling and effective distribution functions for the lead electrons.  The
conductance properties of the latter can be derived with standard methods.
Moreover, for the case of a time-dependent gating voltage, we have revealed
the limitations of the Tien-Gordon approach: While such a treatment provides
the correct expression for the current, it neglects interferences of
different Floquet channels.

A detailed investigation of the recently found shot noise suppression
provided a deeper understanding of this effect.  In particular, the
analytical treatment within a high-frequency approximation can explain the
characteristic emergence of the current suppressions which are accompanied
by a noise maximum and two remarkably low minima.  A numerical study fully
confirmed the analytical results.
For lower driving frequencies, i.e., beyond the high-frequency limit, the
current suppressions become considerably less pronounced.  By contrast, the
shot noise suppression turned out to be more stable.  Thus, since the
current stays remarkably large while the noise is controllable, this regime
is particularly promising for applications.
At first sight, in the limit of strong wire-lead coupling these phenomena
appear quite different.  A closer look, however, revealed that the strong
coupling entails a hybridization of the first and the last site with the
respective lead.  Therefore, the wire behaves qualitatively like a weakly
coupled wire with two sites less.
Moreover, we have found that the noise suppressions are quite sensitive to
an internal bias.  Once the on-site energies of neighboring sites have
differences of the order of the wire-lead coupling energy, the minima of
the Fano factor vanish.
A most interesting application of these results is the development of
current sources with a controllable noise level.

\begin{acknowledgments}

We thank Gert-Ludwig Ingold, J\"org Lehmann, and Michael Strass for helpful
discussions.  This work has been supported by the Volkswagen-Stiftung under
Grant No.\ I/77 217, a Marie Curie fellowship of the European community
program IHP under contract No.\ HPMF-CT-2001-01416 (S.C.), and the DFG
Sonderforschungsbe\-reich 486.

\end{acknowledgments}

\appendix
\section{ac transport voltage}
\label{app:ac.voltage}

Within this work, we focus on models where the driving enters solely by
means of time-dependent matrix elements of the wire Hamiltonian while the
leads and the wire-lead couplings remain time-independent.  An \textit{a
priori} different type of driving is the application of a time-dependent
transport voltage.  In this appendix, we demonstrate that a setup with an
oscillating transport voltage can be mapped by a gauge transformation to a
Hamiltonian of the form \eqref{eq:H}.  Consequently, it is possible to apply the
formalism derived derived in Sec.~\ref{sec:scattering} also to situations
with an oscillating transport voltage.

We restrict the discussion to the case where the electron energies of only
the left lead are modified by an external $\mathcal{T}$-periodic voltage
$V_\mathrm{ac}(t)$ with zero time-average, thus in the left lead
\begin{equation}
\epsilon_{q} \to \epsilon_{q} - eV_\mathrm{ac}(t) .
\end{equation}
The generalization to a situation where also the levels in the right lead
are $\mathcal{T}$-periodically time-dependent, is straightforward.
Since an externally applied voltage causes a potential drop along the wire
\cite{Nitzan2002a, Pleutin2003a, Liang2004a}, we have to assume for
consistency that for an ac voltage, the wire Hamiltonian also obeys a
time-dependence.  Ignoring such a time-dependent potential profile enables
a treatment of the transport problem within the approach of
Refs.~~\onlinecite{Tucker1979a} and ~\onlinecite{Tucker1985a}.  In the
general case, however, we have to resort to the approach put forward with
this work.

We start out by a gauge transformation of the Hamiltonian \eqref{eq:H} with
the unitary operator
\begin{equation}
\label{eq:Uac}
U_\mathrm{ac}(t)
= \exp\Big\{-i\phi(t)\Big(c_1^\dagger c_1^{\phantom{\dagger}}
  + \sum_q c_{Lq}^\dagger c_{Lq}^{\phantom{\dagger}} \Big)\Big\}
\end{equation}
where
\begin{equation}
\phi(t) = -\frac{e}{\hbar}\int^t dt'\,V_\mathrm{ac}(t')
\end{equation}
describes the phase accumulated from the oscillating voltage.  The
transformation \eqref{eq:Uac} has been constructed such that the new
Hamiltonian $\widetilde H(t)=U_\mathrm{ac}^\dagger H(t) U_\mathrm{ac}
-i\hbar U_\mathrm{ac}^\dagger \dot U_\mathrm{ac}^{\phantom{\dagger}}$
possesses a time-independent tunnel coupling.  Since, the operator $c_1$
transforms as $c_1 \to c_1\exp(-i\phi(t))$, the matrix elements
$H_{nn'}(t)$ of the wire Hamiltonian acquire an additional time-dependence,
\begin{equation}
\label{eq:Hw,mod}
\begin{split}
H_{nn'}(t) \to \widetilde H_{nn'}(t)
={} & H_{nn'}(t)e^{-i\phi(t)(\delta_{n'1}-\delta_{n1})} \\
  & + eV_\mathrm{ac}(t) \delta_{n1}\delta_{n'1} .
\end{split}
\end{equation}
The second term in the Hamiltonian \eqref{eq:Hw,mod} stems from $-i\hbar
U_\mathrm{ac}^\dagger \dot U_\mathrm{ac}^{\phantom{\dagger}}$.  Owing to
the zero time-average of the voltage $V_\mathrm{ac}(t)$, the phase
$\phi(t)$ is $\mathcal{T}$-periodic.  Therefore, the transformed wire
Hamiltonian is also $\mathcal{T}$-periodic while the
contact and the lead contributions are time-independent, thus,
$\widetilde H(t)$ is of form~\eqref{eq:H}.

\section{Alternative derivation}
\label{appendix:xi}

In Ref.~~\onlinecite{Camalet2003a}, the expressions \eqref{eq:Ife} and
\eqref{eq:Sfe} for the current and the noise in the wide-band limit have
been derived by eliminating the leads in favor of a stochastic operator.
In this appendix, we detail this approach.
Like in Section \ref{sec:scattering}, we start here also from the
Heisenberg equations \eqref{eq:Heisenberg:lead}--\eqref{eq:Heisenberg:n}
for the annihilation operators.  The ones for the lead operators,
Eq.~\eqref{eq:Heisenberg:lead}, are easily integrated to read
\begin{equation}
\label{eq:c:lead}
c_{Lq}(t)=c_{Lq}(t_0)e^{-i\epsilon_q(t-t_0)/\hbar}
-\frac{iV_{Lq}}{\hbar}\!\!\int\limits_0^{t-t_0}\!\!\! d\tau\,
e^{-i\epsilon_q\tau/\hbar} c_1(t-\tau)
\end{equation}
and $c_{Rq}(t)$ accordingly.  Inserting \eqref{eq:c:lead} into the
Heisenberg equations \eqref{eq:Heisenberg:1,N} for the wire operators
yields
\begin{align}
\dot c_{1/N} =& -\frac{i}{\hbar} \sum_{n'} H_{1/N,n'}(t)\, c_{n'}
             -\frac{\Gamma_{L/R}}{2\hbar}c_{1/N} + \xi_{L/R}(t),\nonumber\\
\dot c_n =& -\frac{i}{\hbar} \sum_{n'} H_{nn'}(t)\, c_{n'}\,,
\quad n=2,\ldots,N-1.
\label{eq:c:xi}
\end{align}
Owing to the wide-band limit, the dissipative terms are memory free.
Within the chosen grand canonical ensembles the operator-valued Gaussian
noise
$\xi_{\ell}(t)=-(i/\hbar)\sum_q
V^*_{\ell q}e^{-i\epsilon_q(t-t_0)/\hbar}c_{\ell q}(t_0)$
obeys
\begin{align}
\label{eq:xi}
\langle\xi_\ell(t)\rangle =  {}& 0,
\\
\label{eq:xi2}
\langle\xi^\dagger_\ell(t)\,\xi_{\ell'}(t')\rangle
 = {}& \delta_{\ell\ell'}\frac{\Gamma_\ell}{2\pi\hbar^2}
\int\!\! d\epsilon\, e^{i\epsilon(t-t')/\hbar}f_\ell(\epsilon) \, .
\end{align} 
The current operator then assumes the form
\begin{equation}
I_L(t)=\frac{e}{\hbar}\Gamma_Lc_1^\dagger(t)c_1(t)
       -e\big\{c_1^\dagger(t)\xi_L(t)+\xi_L^\dagger(t)c_1(t)\big\}.
\end{equation} 

The homogeneous set of equations that corresponds to \eqref{eq:c:xi}
coincides with the equations of motion \eqref{eq:dotUnn} and
\eqref{eq:dotU1nWBL} which are solved by the Floquet states
$|u_\alpha(t)\rangle$.  Thus, the Floquet states $|u_\alpha(t)\rangle$
together with the adjoint states $|u_\alpha^+(t)\rangle$, allow to write
the solution of \eqref{eq:c:xi} in closed form.  In the asymptotic limit
$t_0\to -\infty$, it reads
\begin{equation}
\label{eq:c(t)}
\begin{split}
c_n(t)=
& \int_{0}^{\infty} d\tau\,
  \langle n| U(t,t-\tau) 
\\
& \times\big\{
  |1\rangle\xi_L(t-\tau)+|N\rangle\xi_R(t-\tau)\big\} \, .
\end{split}
\end{equation}
where $U(t,t-\tau)$ is the propagator \eqref{eq:Gtt} for the wire
electrons.

To obtain the current $\langle I_L(t)\rangle$, we insert the operator
\eqref{eq:c(t)} into the expression \eqref{eq:I(t)} and use the expectation
values \eqref{eq:xi2}.  With the Green function \eqref{eq:Gtepsilon}, we
find the still unsymmetric expression
\begin{equation}
\label{eq:Iunsym}
\begin{split}
\langle I_L(t)\rangle
=    \frac{e\Gamma_L}{2\pi\hbar}\int\! d\epsilon\, \Big\{
   & \Gamma_L |G_{11}(t,\epsilon)|^2 f_L(\epsilon) \\
 + & \Gamma_R |G_{1N}(t,\epsilon)|^2 f_R(\epsilon) \\
 + & i[G_{11}^*(t,\epsilon)-G_{11}(t,\epsilon) ] f_L(\epsilon) \Big\} .
\end{split}
\end{equation}
For a symmetrization, we eliminate the backscattering terms, i.e., terms
containing $G_{11}$, by use of the relation \cite{Datta1992a}
\begin{equation}
\label{eq:datta:driven}
\begin{split}
G^\dagger(t,\epsilon)-G(t,\epsilon)
={} &  i\hbar \frac{d}{dt}G^\dagger(t,\epsilon) G(t,\epsilon)
\\
&  + 2 i G^\dagger(t,\epsilon)\Sigma G(t,\epsilon)
\end{split}
\end{equation}
which follows readily from the Floquet representation \eqref{eq:Gtt} of the
propagator and the Floquet eigenvalue equation \eqref{eq:Fs} together with
its adjoint.  A subsequent Fourier transformation with respect to
$\tau=t-t'$ yields Eq.~\eqref{eq:datta:driven}.  By inserting the matrix
element $\langle 1|\ldots|1\rangle$, we obtain from \eqref{eq:Iunsym} for
the time-dependent current the symmetric expression \eqref{eq:I(t)}.

To derive an expression for the zero frequency noise, we insert the
operator \eqref{eq:c(t)} into the definition \eqref{eq:S} of the
current-current correlation function and integrate over the times $\tau$
and $t$.  Again, we employ the relation \eqref{eq:datta:driven} to bring
$S$ into the symmetric form \eqref{eq:Sfe}.



\begin{thebibliography}{10}
\bibitem{on_periodicity}
The $\mathcal{T}$-periodicity of $S(t,t')$ times requires that both $t$ and
  $t'$ are sufficiently far from the preparation time $t_0$. This is ensured by
  considering the limit $t_0\to -\infty$.
\bibitem{on_factor2}
Historically, the zero-frequency noise \eqref{eq:barS} contains a factor $2$,
  i.e.\ $\bar S'=2\bar S$, resulting from a different definition of the Fourier
  transform. Then, the Fano factor is defined as $F=\bar S'/2e|\bar I|$.
\end{thebibliography}
\end{document}